\documentclass[12pt,preprint]{aastex}
\usepackage{graphicx}
\usepackage{psfig}

\shorttitle{X-ray emission of the HD~104237 young stellar group}
\shortauthors{Testa et al.}
\usepackage{color} 
\usepackage{amsmath}           
\usepackage{amsfonts}

\def \deg {$^\circ$}
\def \arcsec {\hbox{$^{\prime\prime}$}}
\def \hd    {HD~104237}
\def \msol {$M_{\odot}$}
\def \rsol {$R_{\odot}$}
\def \mstar {$M_{\star}$}
\def \rstar {$R_{\star}$}
\def \lx   {$L_{\rm X}$}
\def \lxs  {$L_{\rm s}$}
\def \lxt  {$L_{\rm t}$}
\def \fx   {$f_{\rm X}$}
\def \lbol {$L_{\rm bol}$}
\def \nh   {$N_{\rm H}$}
\def \ne   {$n_{\rm e}$}

\def \FHa  {FLG03}

\def \cha   {{\em Chandra}}
\def \hetgs {{\sc hetgs}}
\def \hetg  {{\sc hetg}}
\def \heg  {{\sc heg}}
\def \meg  {{\sc meg}}
\def \xmm      {{XMM-\em Newton}}
\def \acis     {{\sc acis}}
\def \acisi    {{\sc acis-i}}

\def \fexvii  {Fe\,{\sc xvii}}
\def \fexviii {Fe\,{\sc xviii}}
\def \fexix   {Fe\,{\sc xix}}
\def \fexx    {Fe\,{\sc xx}}
\def \fexxi   {Fe\,{\sc xxi}}
\def \fexxii  {Fe\,{\sc xxii}}
\def \fexxiii {Fe\,{\sc xxiii}}
\def \fexxiv  {Fe\,{\sc xxiv}}
\def \fexxv   {Fe\,{\sc xxv}}
\def \arxvii  {Ar\,{\sc xvii}}
\def \arxviii {Ar\,{\sc xviii}}
\def \sxv     {S\,{\sc xv}}
\def \sxvi    {S\,{\sc xvi}}
\def \sixiii  {Si\,{\sc xiii}}
\def \sixiv   {Si\,{\sc xiv}}
\def \mgxi    {Mg\,{\sc xi}}
\def \mgxii   {Mg\,{\sc xii}}
\def \alxii   {Al\,{\sc xii}} 
\def \alxiii  {Al\,{\sc xiii}} 
\def \nixix   {Ni\,{\sc xix}}  
\def \caxvi   {Ca\,{\sc xvi}}
\def \caxviii {Ca\,{\sc xviii}}

\def \neix    {Ne\,{\sc ix}}
\def \nex     {Ne\,{\sc x}}
\def \ovii    {O\,{\sc vii}}
\def \oviii   {O\,{\sc viii}}
\def \nvii    {N\,{\sc vii}}

\newcounter{ion} 
\newcommand{\eli}[2]{\setcounter{ion}{#2}#1{~\sc\roman{ion}}}

\begin{document}
\title{X-ray Emission from Young Stellar Objects in the 
	$\epsilon$ Chamaeleontis Group: the Herbig Ae Star
	\hd\ and Associated Low-Mass Stars}
\author{Paola Testa\altaffilmark{1,2}, David P.\ Huenemoerder\altaffilmark{1}, 
Norbert S.\ Schulz\altaffilmark{1}, Kazunori Ishibashi\altaffilmark{1,3}}
\altaffiltext{1}{Massachusetts Institute of Technology, Kavli 
	Institute for Astrophysics and Space Research, 70 Vassar street, 
	Cambridge, MA 02139, USA; testa@space.mit.edu}
\altaffiltext{2}{Current address: Smithsonian Astrophysical Observatory,
	60 Garden street, MS 58, Cambridge, MA 02138, USA; 
	ptesta@cfa.harvard.edu}
\altaffiltext{3}{NorthWest Research Associates Inc., 3380 Mitchell
	Lane, Boulder, CO 80301, USA}

\begin{abstract}
We present \cha-\hetgs\ observations of the Herbig Ae star \hd\ and 
the associated young stars comprising lower mass stars, in the 
0.15-1.75\msol\ mass range, in their pre-main sequence phase.
The brightest X-ray source in the association is the central 
system harboring the Herbig Ae primary, and a K3 companion.
Its X-ray variability indicates modulation possibly on time scales 
of the rotation period of the Herbig Ae star, and this would imply
that the primary significantly contributes to the 
overall emission. The spectrum of the Herbig Ae+K3 system shows
a soft component significantly more pronounced than in other K-type
young stars. This soft emission
is reminiscent of the unusually soft spectra observed for the single
Herbig Ae stars HD~163296 and AB~Aur, and therefore we tentatively
attribute it to the Herbig Ae of the binary system.
The \hetgs\ spectrum shows strong emission lines corresponding to a
wide range of plasma temperatures. The He-like triplet of \mgxi\ and 
\neix\ suggest the presence of plasma at densities of about 
$10^{12}$~cm$^{-3}$, possibly indicating accretion related
X-ray production mechanism.

The analysis of the zero-order spectra of the other sources
indicates X-ray emission characteristics typical of pre-main
sequence stars of similar spectral type, with the exception of
the T~Tauri \hd-D, whose extremely soft emission is very similar to
the emission of the classical T~Tauri star TW~Hya, and suggests
X-ray production by shocked accreting plasma.
\end{abstract}


\section{Introduction}
\label{s:intro}

Nearby young associations provide ideal targets for the study of star
and planet formation, and have sparked growing interest in the 
past few years.
The early evolution of young stars is strongly dependent on stellar 
mass (e.g., \citealt{Larson72}).
For low-mass stars ($M_{\star}\lesssim 2$\msol) the accretion 
timescales are shorter than the evolution time to the zero-age main
sequence, and these stars can be observed in their pre-main sequence 
(PMS) T Tauri phase \citep{Dantona94}.
In massive protostars ($M_{\star}\gtrsim 10$\msol), however, core 
hydrogen burning already starts during the accretion phase 
\citep{Appenzeller94,Bernasconi96}.  The behavior at the boundaries 
between high and low mass ranges is yet unclear \citep{Palla93}.

Herbig Ae stars (HAe; \citealt{Herbig60}) are PMS stars of 
intermediate mass ($\sim$2--10\msol), and share several 
characteristics with T Tauri stars (TTS), such as IR excess emission, 
irregular photometric variability, and disk properties 
\citep{Waters98,Mannings97}.  
HAe stars show evidence for a disk-like geometry of the 
circumstellar material, and the disk properties are similar to those 
of TTS \citep{Mannings97,Grady99}. 
There is some evidence of magnetic accretion in HAe stars, analogous
to the magnetospheric accretion scenario generally accepted for 
classical TTS (CTTS, i.e.\ TTS that are still actively accreting) and 
are therefore thought to be more massive analogs of CTTS
(e.g., \citealt{Muzerolle04,Grady04,Guimaraes06}). 
While magnetic fields have been detected in a few HAe stars 
(e.g., \citealt{Donati97,Hubrig04}), their origins are still unknown;
however, a shear magnetic model does allow generation of magnetic
fields in the early evolutionary stages of A-stars \citep{Tout95}.

Systematic X-ray studies have shown that HAe stars are moderately 
bright X-ray sources 
\citep{Damiani94,Zinnecker94,Hamaguchi05,Stelzer06}. 
Since main sequence A-stars are not known to be luminous X-ray sources 
due to lack of strong enough winds or magnetic-dynamo driven coronae, 
the findings may imply that the physical characteristics of young 
A-stars stars differ from those of main sequence A-stars. 

HAe X-ray emission mechanisms remain unknown. 
Early studies based on ROSAT and {\em Einstein} observations proposed 
the standard wind shock model as for early-type massive stars 
\citep{Damiani94,Zinnecker94}. However, typical \lx/\lbol\ ratios 
found for HAe stars are significantly higher than the empirical 
luminosity ratio of $10^{-7}$ for early type stars \citep{Skinner04}, 
casting some doubt for this paradigm. Also, the high X-ray temperature 
($> 10^7$~K) component found for a large fraction of HAe X-ray sources 
(see e.g., \citealt{Skinner04,Hamaguchi05,Stelzer06}) cannot be produced 
through this mechanism. 
Others works attribute the emission to magnetic activity (either 
coronal activity as in late-type stars, or magnetic activity due to 
star-disk interaction), or to unresolved late-type companions 
\citep{Hamaguchi05,Skinner96,Skinner04,Stelzer06}.

Another emission mechanism possibly at work in HAe stars is X-ray 
emission from shocks in accreting plasma, proposed as the production 
mechanism for soft X-rays in several CTTS (TW~Hya, BP~Tau, V4046~Sgr, 
MP~Mus; \citealt{Kastner02,Stelzer04,Schmitt05,Gunther06,Argiroffi07}).
For these CTTS, \cha\ and \xmm\ high resolution spectra have revealed
peculiar characteristics with respect to all other observed stellar 
spectra, in particular showing 
(i) soft excess---prominent cool component at temperatures of 
$\sim 3 \times 10^6$~K, TW~Hya being an extreme case, where this soft 
component largely dominates the X-ray spectrum; 
(ii) unusually high density (\ne$\gtrsim 10^{12}$~cm$^{-3}$) for the 
cool ($T \sim 2-4 \times 10^6$~K) plasma, as diagnosed through the 
analysis of the He-like triplets line ratios; 
(iii) peculiar abundances---extremely high Ne, low metal abundances 
\citep{Kastner02,Stelzer04,Drake05}.
Analogously, \cha\ imaging spectra of the HAe star HD~163296 revealed 
a very soft spectrum ($\sim 0.5$~keV) possibly suggesting accretion 
dominated X-ray emission  \citep{Swartz05}, as for the CTTS TW~Hya 
\citep{Kastner02}; \cite{Grady07} also find extremely soft X-ray emission
($\sim 0.25$~keV) from the Herbig Ae star HD~169142.
Recently, \xmm\ observations of the HAe star AB~Aur have shown a very 
similar soft spectrum \citep{Telleschi07}. In this case, the 
high resolution spectrum provides additional diagnostics: the high 
forbidden/intercombination ratios ($f/i$) of the He-like triplets, 
especially \ovii, indicate the absence of high density and high UV 
field, expected if X-rays are produced in the accretion shocks, and 
rule out accretion-related X-ray emission for AB~Aur. 
For AB~Aur, which has spectral type A0, the photospheric UV emission 
is strong enough to alter the O He-like triplet up to several stellar 
radii from the surface; the large measured \ovii\ $f/i$ line ratio, 
constrains the emission to originate high above the photosphere
($\gtrsim 3$\rstar) and therefore it implies that also coronal 
emission is unlikely for AB~Aur.
A plausible alternative for this source is X-ray emission from 
magnetically confined winds, as in the scenario of \cite{Babel97}.
It is important to keep in mind that some of the different X-ray 
emission mechanisms might not be mutually exclusive and instead they 
might coexist, as seems to be the case for some CTTS 
\citep{Schmitt05,Gunther06,Argiroffi07} where both magnetically 
confined coronal plasma and accretion streams likely contributes to 
the overall X-ray emission.

\hd\ is one of the nearest known ($d=116$~pc; \citealt{Perryman97}) 
and well studied HAe stars, and it is also a strong X-ray source 
(\lx~$ \sim 2 \times 10^{30}$~erg~s$^{-1}$).
\cha\ observations have spatially resolved the X-ray emission from 
\hd, showing that the brightest X-ray source is coincident with the 
HAe source, and identifying 4 previously unknown nearby sources, 
likely TTS forming a young stellar group associated with \hd\ 
(\citealt{Feigelson03}, hereafter \FHa; see \S\ref{s:targets}).
\cha\ (\FHa) and \xmm\ \citep{Skinner04} imaging spectra
of this HAe star have shown the presence of hot plasma 
($\gtrsim 2$~keV), taken as evidence of magnetic activity. 
\hd\ with its proximity, high \lx, and low line-of-sight extinction 
is a unique target to study the X-ray emission mechanism in HAe stars,
through high resolution spectroscopy.  
High resolution spectra provide plasma diagnostics which are the only
means for detailed study of the physical conditions of the emitting
material.  Cool, dense plasma ($\log T\mathrm{[K]}\sim 6.5$, 
$\log n_e [\mathrm{cm^{-3}}] > 12$) is now thought to be a signature 
of accretion, while variable hot emission is indicative of magnetic 
activity.  

In this paper we present \cha-\hetgs\ observations of \hd. The 
relatively long exposure of $\sim 145$~ks provides detailed 
information on the X-ray variability of the HAe star, as well as of 
other stars in the field, and it yielded a well exposed \hetgs\ 
spectrum of the HAe star providing detailed diagnostics for the 
emitting plasma.

We present the characteristics of the targets in \S\ref{s:targets}.
The observations and our techniques of line flux measurement and 
spectral analysis are briefly described in \S\ref{s:obs}. 
The results are presented in \S\ref{ss:results}.   
We discuss the results and draw our conclusions in \S\ref{s:discuss},
and \ref{s:conclusions}.

\section{\hd: the Herbig Ae star and the T Tauri association}
\label{s:targets}

\hd\ is the optically brightest HAe star, and it is also a moderately 
strong X-ray source with $\log L_{\rm X} \sim 30.5$~erg/s, observed 
with ROSAT \citep{Alcala95}, ASCA \citep{Skinner96}, \cha\ 
(\FHa), and \xmm\ \citep{Skinner04}.

\begin{figure}[!ht]
\centerline{\psfig{figure=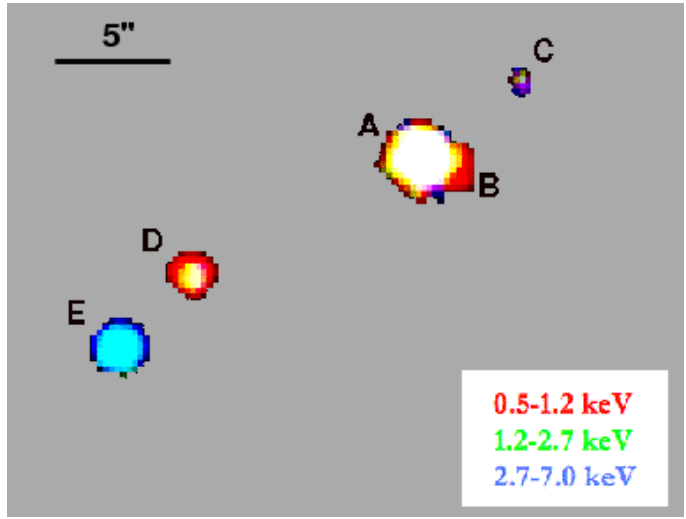,width=9cm}}
\caption{Color coded \cha\ image of the \hd\ field from our 
	$\sim 145$~ks \hetgs\ observation.
	The X-ray sources of the young stellar group are labeled
	following Feigelson et al. (\FHa).
	The separation between sources is listed in 
	Table~\ref{tab:stars}; all sources are within 
	$\sim 20$\arcsec.
	\label{fig:img}}
\end{figure}

\cha\ imaging observations, thanks to the unprecedented sub-arcsecond
spatial resolution, have revealed the presence of four other X-ray
emitting young stellar objects associated with \hd, labeled B--E, A 
being the HAe star (Figure~\ref{fig:img}; see also \FHa). 
\cite{Grady04} (hereafter G04) have conducted intensive high angular resolution 
multiwavelength imaging (optical, NIR, mid-IR), and spatially 
resolved spectroscopy (optical, UV, FUV) to probe the 
characteristics of A, and the associated star-forming environment.
The comprehensive study by G04, has provided detailed information on 
the HAe star and constraints on possible close companions.

The stellar parameters of the members of this young ($\sim$~2-5~Myr) 
stellar group are listed in Table~\ref{tab:stars}.  IR excess, 
H$\alpha$ emission, and Li absorption indicate a T Tauri nature 
for the components B, D, and E (\FHa; G04).  
\FHa\ hypothesized that C is a late M-type TTS or a brown dwarf.

\begin{deluxetable}{lcccc}
\tablecolumns{5} 
\tabletypesize{\footnotesize}
\tablecaption{Characteristics of \hd\ members. \label{tab:stars}}
\tablewidth{0pt}
\tablehead{
 \colhead{Object} & \colhead{Spec.\ Type \tablenotemark{a}} & 
 \colhead{EW(H$\alpha$) \tablenotemark{b}}  &  
 \colhead{offset \tablenotemark{c}}   &  \colhead{comments} 
}
\startdata 
A \tablenotemark{d} & A7.5Ve-A8Ve  & -24.5 & ... & HAe, IR and UV excess, jets \\
B & M3-4              & ...   & $1.365$\arcsec$ \pm 0.019$  & IR excess 		  \\
C & ...               & ...   & $5.275$\arcsec$ \pm 0.033$  & BD (\FHa) or M6V (G04)	  \\
D & M2-3              & -9    & $10$\arcsec$.72 \pm 0.05$   & H$\alpha$ near CTTS limit   \\
E & K3IVe             & -9    & $14$\arcsec$.88 \pm 0.05$   & H$\alpha$ near CTTS limit, IR excess  \\
\enddata 
\tablenotetext{a}{Ranges of spectral types from \citet{Feigelson03,Grady04,Luhman04}.}
\tablenotetext{b}{Equivalent width, in \AA, of H$\alpha$ emission 
        from \citet{Acke05,Grady04}.}
\tablenotetext{c}{from \cite{Grady04}.}
\tablenotetext{d}{A has a K3 spectroscopic companion within $\sim 0.15$
	AU \citep{Bohm04} whose parameters are listed in 
	Table~\ref{tab:Hae}.}
\end{deluxetable}

\begin{deluxetable}{lccl}
\tablecolumns{3} 
\tabletypesize{\footnotesize}
\tablecaption{Characteristics of \hd-A and its spectroscopic
	companion. \label{tab:Hae}}
\tablewidth{0pt}
\tablehead{
 \colhead{Parameter} & \colhead{\hd-A}  & 
 \colhead{companion}  & \colhead{Ref.\ \tablenotemark{a}} 
}
\startdata 
d             & 116~pc        &      &  1  \\
\mstar/\msol    & 2.25          & 1.75 &  2  \\
\rstar/\rsol    & $2.5 \pm 0.2$  &     &  3  \\
$\log (L_{\star}/L_{\odot})$ & $1.42^{+0.04}_{-0.07}$ & 0.5 &  2 \\
$\log (T_{\rm eff})$   & 3.87     & 3.675 &  2   \\
Age                    & 2-5~Myr  &       &  2,5 \\
$v \sin i$             & $12 \pm 2$~km~s$^{-1}$ &  &  4  \\
$i$                    & $18$\deg$^{+14}_{-11}$, $23$\deg$^{+9}_{-8}$ & & 5,3 \\
$P_{\rm rot}$ \tablenotemark{b} & $100 \pm 5$~hrs  & & 3  \\
$P_{\rm orb}$                & 19.859~days     & &  2  \\
$e$ \tablenotemark{c}           & 0.66486         & &  2  \\
\enddata 
\tablenotetext{a}{References: (1) \cite{Perryman97}; (2) \cite{Bohm04}; 
	(3) \cite{Bohm06}; (4) \cite{Donati97}; (5) \cite{Grady04}. }
\tablenotetext{b}{Periodicity of modulation observed in H$\alpha$, and possibly
	associated to rotational modulation by \cite{Bohm06}. }
\tablenotetext{c}{Eccentricity of the binary orbit. }
\end{deluxetable}

\vspace{-0.6cm}
\paragraph{\hd-A:} 
\hd-A has been identified as an Herbig Ae star by \cite{Hu89},
based on its optical spectrum and IR excess.
It was initially assigned a spectral type A0pe by \cite{Walker88},
then A4\,{\sc v}e by \cite{Hu91}, and recently G04 have revised its 
classification as A7.5\,{\sc v}e-A8\,{\sc v}e.
The HAe star presents evidence of activity--as indicated by line 
emission, UV excess, outflows--and is still accreting 
at an estimated rate of $\sim 10^{-8}$ \msol/yr (G04, 
\citealt{GarciaLopez06}). 
UV high resolution spectra show P Cygni profiles for several lines
\citep{Blondel06}.
G04 also found a bipolar microjet and Herbig-Haro knots; from the 
analysis of proper motion of a knot they derived the inclination 
of the system to be $18$\deg$^{+14}_{-11}$.  
The inclination may be a meaningful parameter for the
detection of accretion signatures, assuming the geometry of
magnetospheric accretion models; for instance, the
CTTS TW~Hydrae and BP~Tau, whose soft X-ray emission appears 
accretion dominated, have both nearly pole-on aspects of 10\deg\  
and 30\deg\ respectively. 
The observations suggest a magnetospheric accretion scenario for \hd-A
similar to CTTS (G04). A magnetic field of this HAe has been marginally
detected ($\sim 50$~G; \citealt{Donati97}).
If the geometry of the accretion is similar in \hd-A we will likely
observe almost along the accretion stream.
The \hd-A disk extends at most to 0.6\arcsec\ ($\sim$ 70~AU). 

{\em HST} observations put tight constraints on the distance of 
any unknown close companion to A (other than the resolved 
X-ray source B at $\sim 1.4$\arcsec\ separation; G04), 
excluding the presence of a companion with distance $r$ 
to the A star 0.05\arcsec$<r<$1\arcsec, implying that a possible
companion would have to be closer than $\sim 5$~AU to A.
\cite{Bohm04} through high resolution spectral observations 
(aimed at the study of $\delta$~Scuti pulsation observed for
this HAe star) revealed the presence of a spectroscopic binary
companion to the HAe star, and determined the orbital parameters.
They find a mass ratio $M_{\rm primary}/M_{\rm secondary} = 1.29 \pm 0.02$,
and an orbital period of about 20~days, implying an average separation
of $\sim 0.15$~AU. They assign a spectral type K3 to this spectroscopic
companion which is responsible for the Li-line traces in the spectrum 
of the HAe star revealed by \FHa\ and which they attributed to 
contamination from the adjacent B component. The parameters of the HAe 
star and the K3 close companion are included in Table~\ref{tab:Hae}.
Recently, \cite{Tatulli07} using near-IR interferometric observations
of this HAe star modeled the spatial distribution of the circumstellar 
material and outflows on AU scales; their observations are compatible
with outflowing wind launched in the vicinity of the dusty disk inner 
rim ($\sim 0.2-0.5$~AU). Considering the very small distance between
the spectroscopic binary components, the circumstellar disk of the
HAe star is likely a circumbinary disk, and this could in 
principle affect the magnetospheric accretion scenario and other
phenomena occurring on small scales close to the stars.

The presence of this later type companion, unresolved in the X-ray 
observations, significantly changes the perspective for interpretation
of the X-ray emission of \hd-A, as K-type PMS stars are known strong
X-ray emitters (e.g., \citealt{Getman05}). 
We will thoroughly discuss this issue in \S\ref{s:discuss}.

\vspace{-0.3cm}
\paragraph{\hd-B:} 
The proximity of \hd-B to the HAe star hampers the precise 
determination of its stellar parameters, as thoroughly discussed by 
G04. \FHa\ tentatively assigned a K spectral type to B due to the 
presence of Li-line features in the optical spectrum of \hd-A; 
however, these features are actually produced by the the K-type close
companion subsequently discovered as discussed above.
G04 finds that \hd-B is likely a M-type TTS on the basis of its 
infrared excess, and of its PSF in HST/STIS images, compatible with 
M3-M4 spectral type.

\vspace{-0.3cm}
\paragraph{\hd-C:} 
\FHa\ has tentatively classified \hd-C as a brown dwarf candidate.

\vspace{-0.3cm}
\paragraph{\hd-D:} 
The H$\alpha$ equivalent width, close to the classical TTS limit,
the Li features, and the X-ray luminosity indicate the T Tauri nature
of \hd-D (G04). 

\vspace{-0.3cm}
\paragraph{\hd-E:} This star has a late spectral type, broad
H$\alpha$ emission, and photometric variability, which are typical of
accreting CTTS (G04). This source was also variable in X-rays
and showed a remarkable change in emission level by a factor $\sim 8$
between the two short \acisi\ observations of \FHa, 
who also found it to be much more absorbed than
the other stars in the association. This suggests that the obscuration
is due to local circumstellar material.

\section{Observations and Analysis}
\label{s:obs}

We observed the Herbig Ae star \hd\ and the group of young stellar
objects associated to it with the {\em Chandra} High Energy
Transmission Grating Spectrometer (see \citealt{hetg05} for a
description of the instrumentation) for a total exposure time of
$145$~ks.  The details of the observations are
presented in Table~\ref{tab_obs}.  

We also re-analyzed the \acis\ imaging-mode observations of \FHa\
(ObsID 2404 and 3428).

The data used here have been reprocessed using standard CIAO
\citep{CIAO} v3.4 tools.  Effective areas were calculated using
standard CIAO procedures, which include an appropriate
observation-specific correction for the time-dependent \acis\
contamination layer.

Spectral analysis of grating and zeroth order spectra were done with
the Interactive Spectral Interpretation System (ISIS\footnote{ISIS is
  available at http://space.mit.edu/cxc/isis/}) version 1.4.2
\citep{Houck00}.

In Figure~\ref{fig:img} we present a color coded image of the
\cha-\hetgs\ observations where photons energies in the range
$2.7-7.0$~keV correspond to blue, $1.2-2.7$~keV to green, and
$0.5-1.2$~keV to red. The X-ray sources are labeled following the
classification of \FHa.

\begin{deluxetable}{cccccccccc}
\tablecolumns{10} 
\tabletypesize{\footnotesize}
\tablecaption{Parameters of the HETG observations and count rates of X-ray sources
		\hd-A, B, C, D, E.
		 \label{tab_obs}}
\tablewidth{0pt}
\tablehead{
 \colhead{Obs ID} & \colhead{Start date and time} & \colhead{$t_{\rm exp}$ [ks]}  & 
 \multicolumn{7}{c}{count rate [cts/ks]} \\ \cline{4-9}
 & & & \multicolumn{2}{c}{A} &  &  \colhead{B \tablenotemark{a}} &  \colhead{C \tablenotemark{a}} & 
 	\colhead{D \tablenotemark{a}} &   \colhead{E \tablenotemark{a}} \\ \cline{4-5}
 & & & 0$^{\rm th}$ \tablenotemark{a}  &  1$^{\rm st}$ \tablenotemark{b}  &  &  &  &  
}
\startdata
 7319  &  2006-04-11 23:46:08  &  43.39  & 23.0  &  33.7  &  &  0.67 &  0.21  &  1.5  &  3.7 \\
 7320  &  2006-04-13 10:13:06  &  43.83  & 19.8  &  28.0  &  &  0.62 &  0.73  &  1.4  &  2.3 \\
 6444  &  2006-04-16 09:51:47  &  12.57  & 20.4  &  27.8  &  &  1.85 &  0.48  &  1.9  &  3.9 \\
 7326  &  2006-06-13 05:32:27  &  45.61  & 35.0  &  46.9  &  &  0.85 &  0.31  &  2.2  &  5.1 \\
 \enddata 
\tablenotetext{a}{count rate from zero order counts in the 0.5-8~keV energy range.}
\tablenotetext{b}{count rate from first order (MEG -1,+1, HEG -1,+1) dispersed counts,
			in the 1.5-26\AA\ wavelength range.}
\end{deluxetable}

In the following section we present the analysis of the X-ray emission
of all sources of this young stellar association.  We analyzed the low
resolution spectra and variability for all group members, and the high
resolution spectrum of source A which has enough counts for a
meaningful analysis of the dispersed photons.

\section{Results}
\label{ss:results}
In the total $\sim 145$~ks \hetg\ observations we detected all sources
studied in \FHa.  Source A is the strongest X-ray source but even
for this source the pileup in the zeroth order is essentially 
negligible (up to a few \%).
Source E and D have comparable count rates about an order of magnitude
smaller than for source A.  Source B, which is at an angular distance
of $\gtrsim 1$\arcsec\ (\FHa, G04), is apparent as an elongation of
source A in the SW direction; even though there is a small
contamination of source B in the spectrum of source A its contribution
to the A+B spectrum is lower than 10\%.  Finally, source C is detected
at a $3 \sigma$ level only for a portion of the observing time.

The \hetg\ observations presented here provide significantly better
statistics than the previous \acisi\ observations analyzed by \FHa.
For source A, we obtained more than 3700 integrated zero-order counts
vs.\ 640 of the short \acis\ observations. For the sources B, C, D and
E we obtain 152, 54, 251, and 543 zero-order counts which are a factor
2.5 to 9 times higher than the counts of the \acisi\ observations
which were 44, 6, 70, 225, respectively.

In order to extract  
the zero-order spectrum
of source B we 
needed to take into account the contamination of source A. 
This contamination was estimated by extracting the counts of 
source A in an annular region 2-4 pixels from the center, and 
excluding a sector containing source B (between -0.64 and 0.24~rad, 
where 0.0 is west and positive angles are in the north direction).  
These counts were then scaled to the region of extraction of source 
B and used as a background for this source.

\subsection{Variability}

\begin{figure*}[!tbhp]
\centerline{\psfig{figure=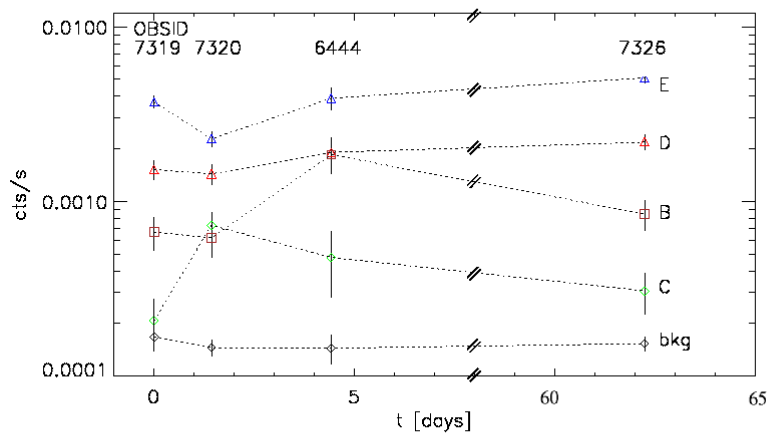,width=11cm}}
\caption{Zero order lightcurves of sources B, C, D, E, and for the 
	background: we plot count rates with error bars for the four 
	observations as a function of time; the last observation, 
	carried out two months apart the first one is shifted by 
	-50~days. The background rate was derived from a circular 
	region close to source C, with a radius 4 times larger 
	than the default extraction radius used for all the sources. 
	The counts for source B are derived as described in the text
	with a procedure we devised to take into account and subtract
	the contamination of source A.
	\label{fig:lc_CDE}}
\end{figure*}

All sources in the \hd\ field appear to be variable.  In
Figure~\ref{fig:lc_CDE} we show the lightcurves obtained from
zero-order photons for sources B, C, D, E.  In order to show how
significant the detection of source C is in the different segments of
observation we also derived a lightcurve for the background, using an
extraction region, close to source C, and with radius 4 times larger
than the default value used for the extraction of the spectra of all
sources, and then scaling the rate to the source region areas.  All
sources present significant variability over time scales of days to
months.

\begin{figure*}[!hr]
\centerline{\psfig{figure=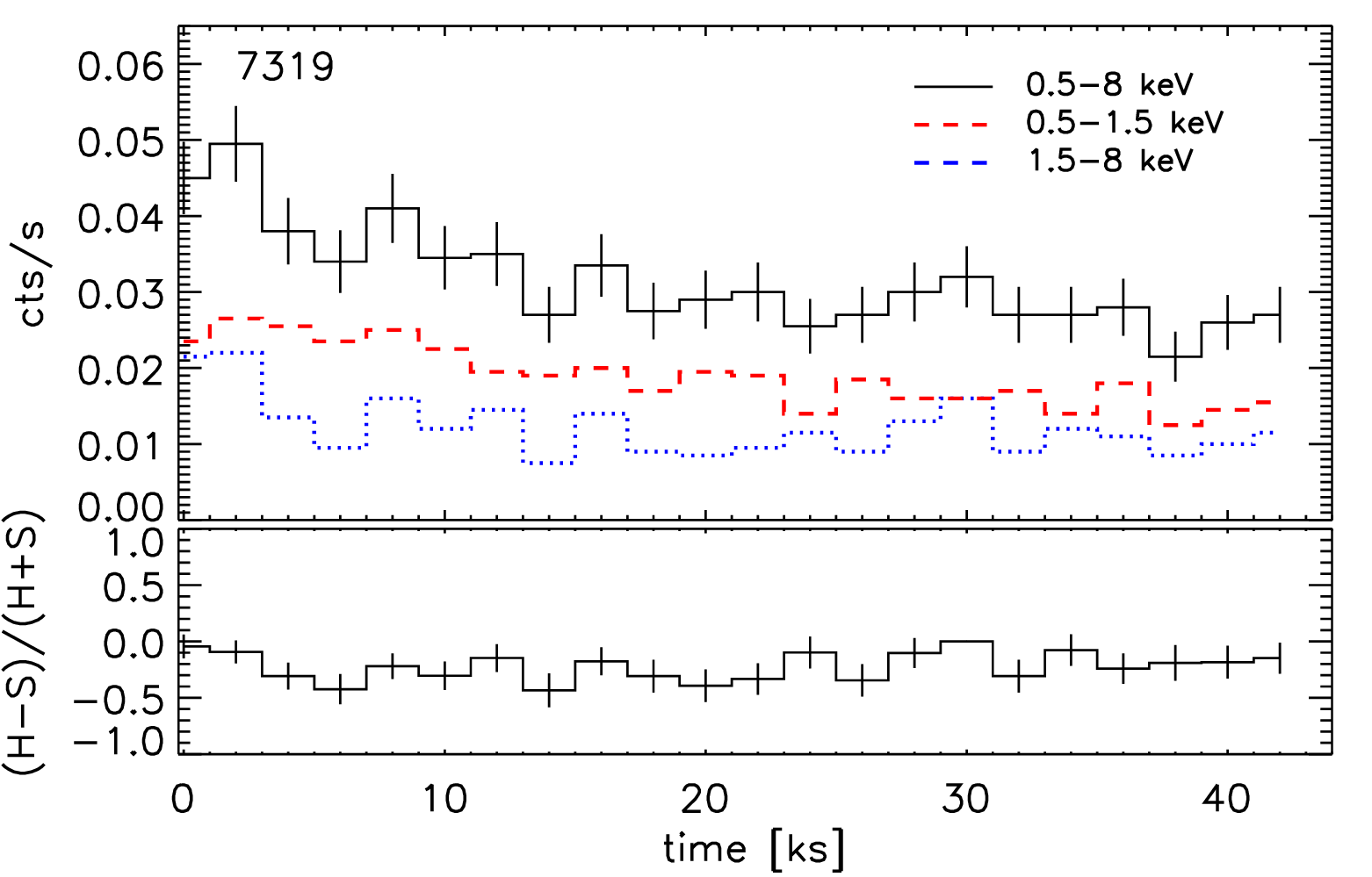,width=8cm}
	    \psfig{figure=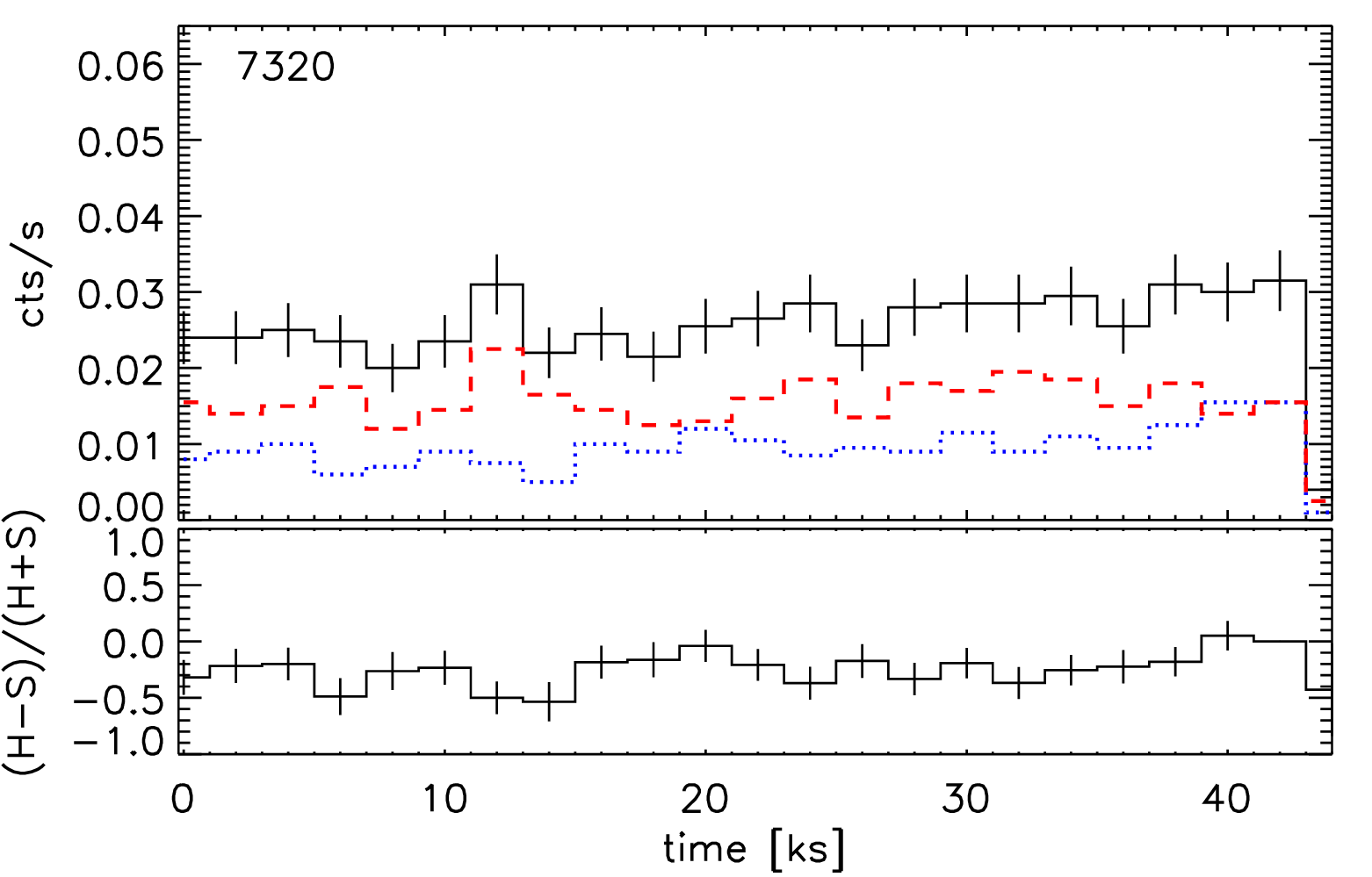,width=8cm}}
\centerline{\psfig{figure=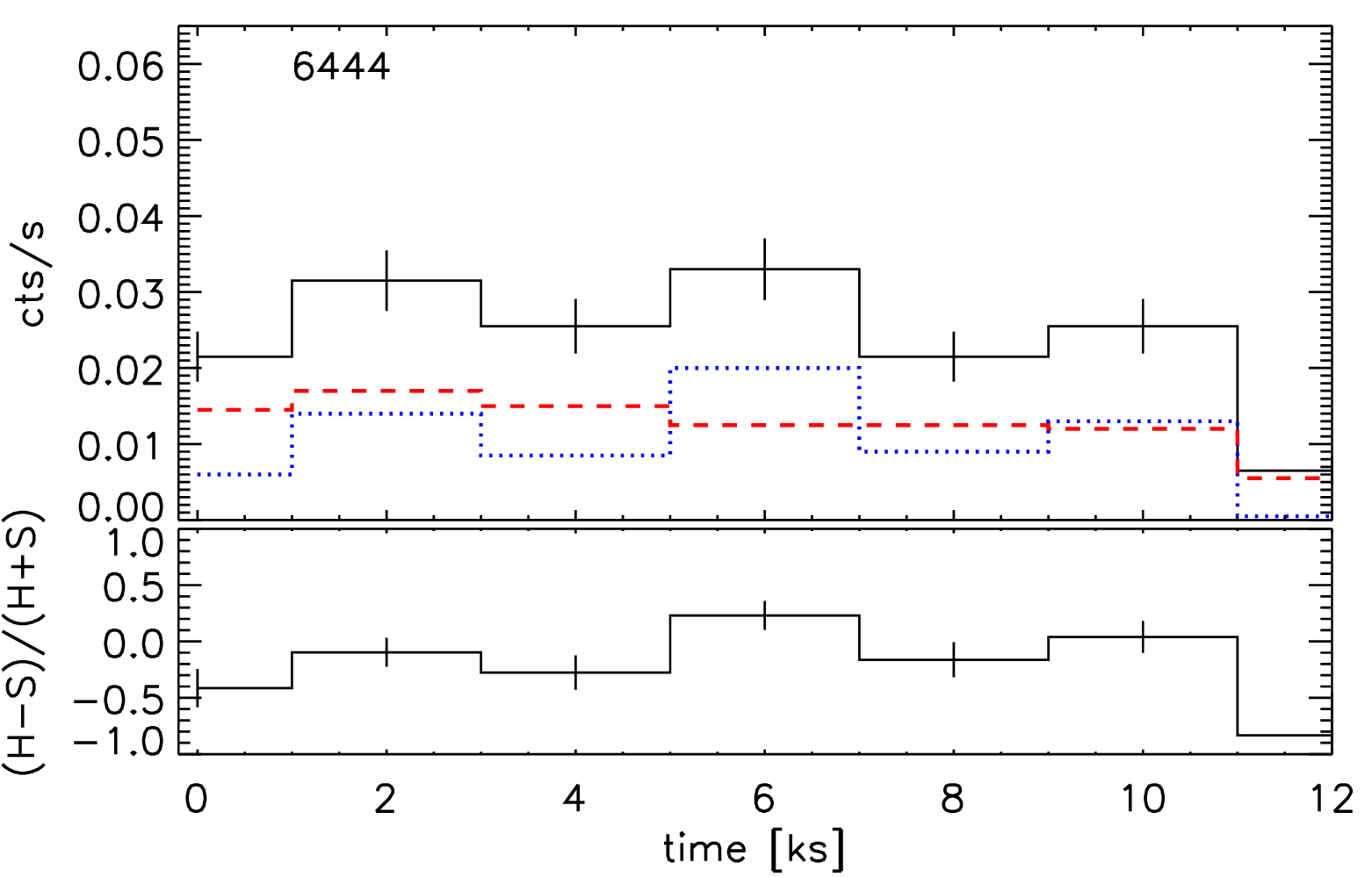,width=8cm}
	    \psfig{figure=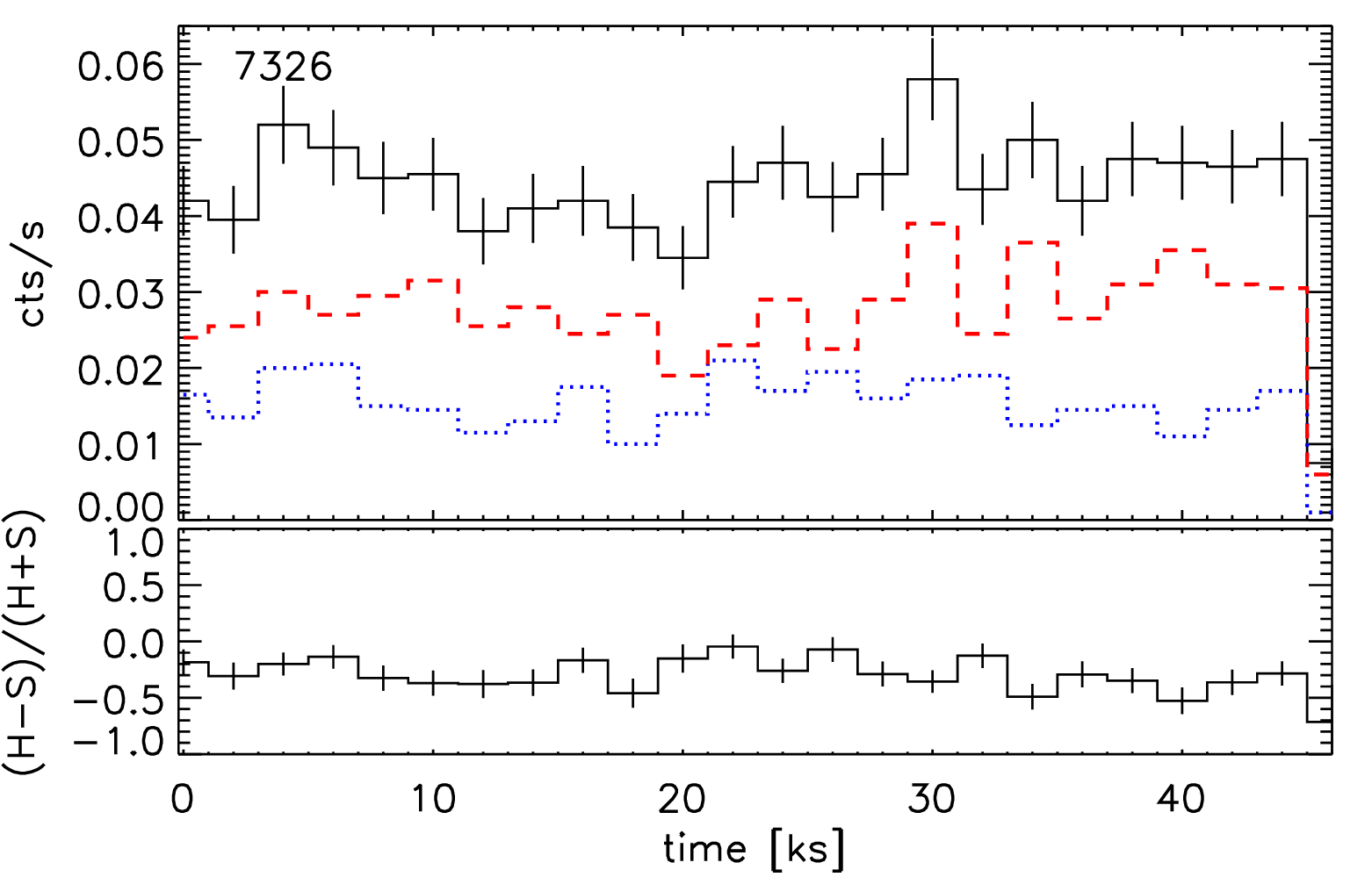,width=8cm}}
\caption{Lightcurves of first order dispersed photons of source A 
	for each of the	four observations, using a binsize of 2~ks. 
	The upper panel of each plot shows the lightcurves of all 
	the first order dispersed photons (solid line with error
	bars), and the lightcurves in two energy bands: a hard band 
	($1.5-8$~keV; dotted line) and a soft band ($0.5-1.5$~keV; 
	dashed line). The lower panels show the corresponding 
	hardness ratio defined as [hard-soft]/[hard+soft].
	\label{fig:lcHETG}}
\end{figure*}

In Figure~\ref{fig:lcHETG} we plot the lightcurves of source A
as obtained from the dispersed photons. In the upper panels we
plot the lightcurve of all dispersed photons (black curves),
and also the lightcurves in two spectral bands in order to show
possible changes in spectral hardness: the hard band corresponding
to the $1.5-8$~keV energy range (blue curves), 
and the soft band corresponding to the $0.5-1.5$~keV range 
(red curves). The lower panels show the corresponding
hardness ratio (defined as [hard-soft]/[hard+soft]).
These plots show that the source presents continuous variability 
(within a factor $\sim 2$ from an average value) on
time scales from hours to days.  However the observed variability
does not show obvious similarities with very dynamic events such as 
flares, which are typical of active stellar coronae and are generally 
characterized by significant increases in hardness ratio and time 
scales of hours up to days for the most extreme events
observed in very active young stars.

We searched for possible periodicity in the lightcurves of source A, 
in particular exploring possible periodic variation on the 
characteristic time scales of the system: the orbital period of the 
binary system ($P_{\rm orb}=19.859$~days; \citealt{Bohm04}), and the 
rotation period. 
The rotation period of this Herbig Ae star is however not well 
determined, therefore we used different estimates. Using the 
$v \sin i = 12$~km~s$^{-1}$ obtained by \cite{Donati97} (derived from 
several photospheric spectral lines), the inclination $i = 18$\deg\ 
estimated by G04, and a stellar radius $R_{\star} = 2.5$\rsol\ 
\citep{Bohm06} we obtain a rotation period $P_{\rm rot} = 3.26$~days; 
however, considering the uncertainties on all 
these parameters\footnote{For instance, \cite{Blondel06} assume very 
different values for both $v \sin i$ (150~km~s$^{-1}$), and $i$ (53\deg), 
which they infer from optical lines, and the redshifted component of 
the Ly$\alpha$ emission respectively.}, the rotation period is poorly 
determined. \cite{Bohm06} present a possible indication of rotational 
modulation by studying modulation of lines forming close to the 
stellar photosphere: they find modulation in the H$\alpha$ line 
that might be due to rotation and it would yield 
$P_{\rm rot} = 100 \pm 5$~hrs, i.e.\ $P_{\rm rot} = 4.17 \pm 0.21$~days.
Therefore we performed a period analysis in this range of periods and
we find a best fit period of 95~hrs.

\begin{figure*}[!hr]
\centerline{\psfig{figure=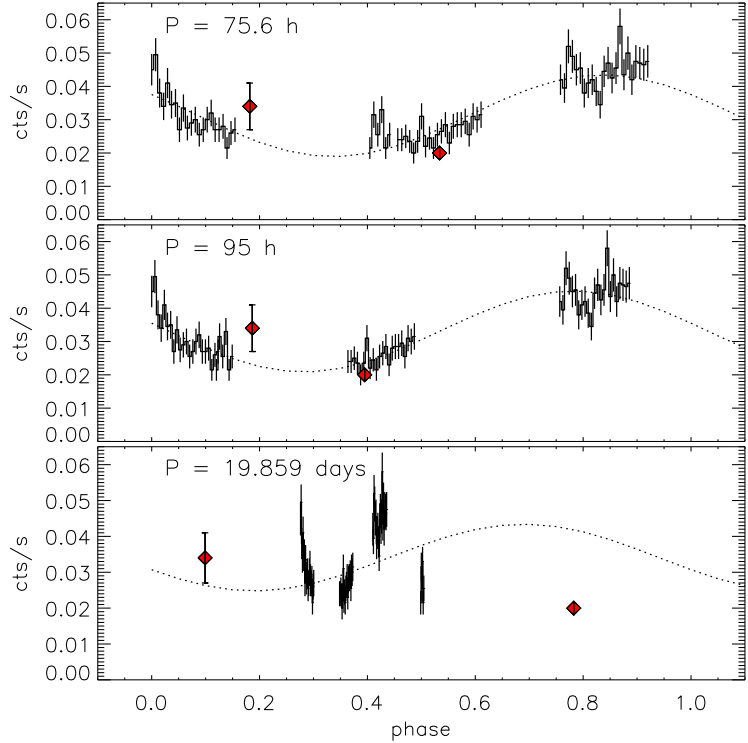,width=10cm}}
\caption{Lightcurves of dispersed photons of source A (binsize of 2~ks) 
	phase-folded assuming three different periods: {\em top --}
	$P = 75.6$~h obtained assuming $v \sin{i} = 12$~km~s$^{-1}$ 
	\citep{Donati97}, $i=18^{\circ}$ (G04), and 
	$R_{\star} = 2.5 R_{\odot}$ \citep{Bohm06};
	{\em middle --} $P = 95$~h (within range of possible rotational
	modulation observed in $H \alpha$ by \citealt{Bohm06}; see text);
	{\em bottom --} $P = 19.859$~days~$= P_{\rm orb}$ of the binary 
	system of the Herbig Ae star \hd\ + K3 companion \citep{Bohm04};
	phase 0 corresponds to the periastron passage. 
	Filled symbols indicate the zero-order count rate predicted 
	from modeling the \acisi\ observations (see text and 
	Table~\ref{tab:fit_A_acis}).
	The dotted line represent a fit to the data with a simple 
	sinusoidal curve plus a constant; while this function does not 
	represent a specific physical model, the constant function 
	can be interpreted as the emission level of the quiescent 
	(i.e.\ more homogeneously distributed) corona, while the
	rotational modulation would be likely caused by the appearing 
	and disappearing from view of compact regions of more intense 
	X-ray emission.	
	\label{fig:lcHETG_phf}}
\end{figure*}

Figure~\ref{fig:lcHETG_phf} shows the phase folded lightcurves
for three different periods. All plots indicate that our X-ray observations
do not provide a large phase coverage even for the shortest period, 
therefore making difficult to identify possible periodic modulation
of the X-ray emission. Nevertheless, the observed lightcurve seem to be 
compatible with modulation on time scale comparable with the rotation
period as estimated by \cite{Bohm06}; if true this would imply that 
the Herbig Ae star is likely a significant contributor to the overall 
X-ray emission of the binary system.
The variability does not show correlation with the orbital period of the 
\hd-A system of 19.859~days (bottom panel of Fig.~\ref{fig:lcHETG_phf}).

In Figure~\ref{fig:lcHETG_phf} we also superimpose the count rates
estimated for the two early \acisi\ observations studied by \FHa\
(ObsID 2404 and 3428).  \FHa\ in their analysis assumed negligible
\nh\ and did not account for pileup. In order to take into account
these effects and estimate the expected count rate for these two
observations, we re-analyzed the data
using the pileup model of \citep{Davis01} as implemented in ISIS.
This model accounts for coincident detection of multiple photons and
has the advantage of using all the counts in the source region, not
just those in an outer, un-piled annulus.
We found a two-temperature model was necessary, and that the pileup
fraction was about 20\%.  In fitting the spectra we fixed the
abundances to the values derived from the \hetgs\ analysis, and the
\nh\ to $2 \times 10^{21}$~cm$^{-2}$ (see detailed discussion in
\S\ref{s:lowres}).  We note that our analysis of the \acisi\ spectra
of \hd-A yields for both temperature components lower values with
respect to those found by \FHa; also, our derived luminosity values
are larger than the values found by \FHa\ (by about 20\% and 50\%
respectively in the two observations). These discrepancies are due to
the inclusion of interstellar absorption and pileup effects.

\begin{deluxetable}{lcc}
\tablecolumns{3} 
\tabletypesize{\footnotesize}
\tablecaption{Parameters and uncertainties of 2-T models 
	fitting the \acisi\ spectra of \hd-A.  \label{tab:fit_A_acis}}
\tablewidth{0pt}
\tablehead{
 \colhead{Parameter \tablenotemark{a}}  &  \colhead{ObsID 2404}  &  \colhead{ObsID 3428}
}
\startdata

 T$_1$     ($10^6$~K)             &  4.7 [4.1-5.3]    &  4.8 [4.1-5.3]    \\
 EM$_1$    ($10^{52}$~cm$^{-3}$)  &  19.6 [17.7-20.9] &  22.9 [13.7-26.6] \\
 T$_2$     ($10^6$~K)             &  36.8 [26.8-56.7] &  18.8 [16.3-21.4] \\
 EM$_2$    ($10^{52}$~cm$^{-3}$)  &  3.4 [2.9-4.2]    &  15.4 [9.9-17.4]  \\
 \lxs\ ($10^{30}$~erg~s$^{-1}$)   &  2.7              &  4.1	          \\
 \lxt\ ($10^{30}$~erg~s$^{-1}$)   &  3.0              &  4.7     	  \\
 \hetgs\ modeled zero order count rate &  0.020       &  0.035            \\
 \acisi\ modeled (unpiled) count rate  &  0.16        &  0.27             \\
 \acisi\ observed count rate           &  0.12        &  0.15             \\
\enddata 
\tablenotetext{a}{The 1$\sigma$ confidence intervals are listed in 
	square brackets. \nh\ is fixed at $2 \times 10^{21}$~cm$^{-2}$ 
	and element abundances are fixed to the values derived from 
	the \hetgs\ spectrum (see Table~\ref{tab:abund}).
	\lxs\ and \lxt\ are the X-ray luminosity values in the 
	0.5-2.0~keV and 0.5-8.0~keV respectively, as in \FHa; these 
	value are corrected for absorption.}
\end{deluxetable}

\subsection{Low resolution spectra}
\label{s:lowres}
The color coded image (Figure~\ref{fig:img}) shows that source E is 
characterized by a hard spectrum, source D is on the contrary extremely 
soft, and source A is an intermediate case.
Here we present the low resolution spectra of source A, B, C, D, and E
and the results of the fits to these spectra with isothermal or 
two-temperature plasma models.

For source A, the higher signal to noise allow us to study separately 
the low resolution spectra of each observation to highlight possible
changes in the spectral characteristics with changes in the X-ray 
emission level. For all other sources due to the poorer statistics 
we combined the spectra of all four observations in order to improve the 
signal-to-noise ratio and better constrain the average spectral parameters.

We extracted the low resolution spectra of source A for each observation,
and performed a fit with a thermal model of plasma in collisional
ionization equilibrium with variable absorption. A single temperature model, 
even with variable abundances, does not provide a good fit to the spectrum. 
Therefore we use a 2-T component model with single \nh.
The fits to the spectra obtained for the separate observations do not 
show significant changes in the spectral parameters within the uncertainties,
therefore we summed up the spectra for all four observations to better
constrain the parameters of the model.

\begin{figure*}[!tbhp]
\centerline{\psfig{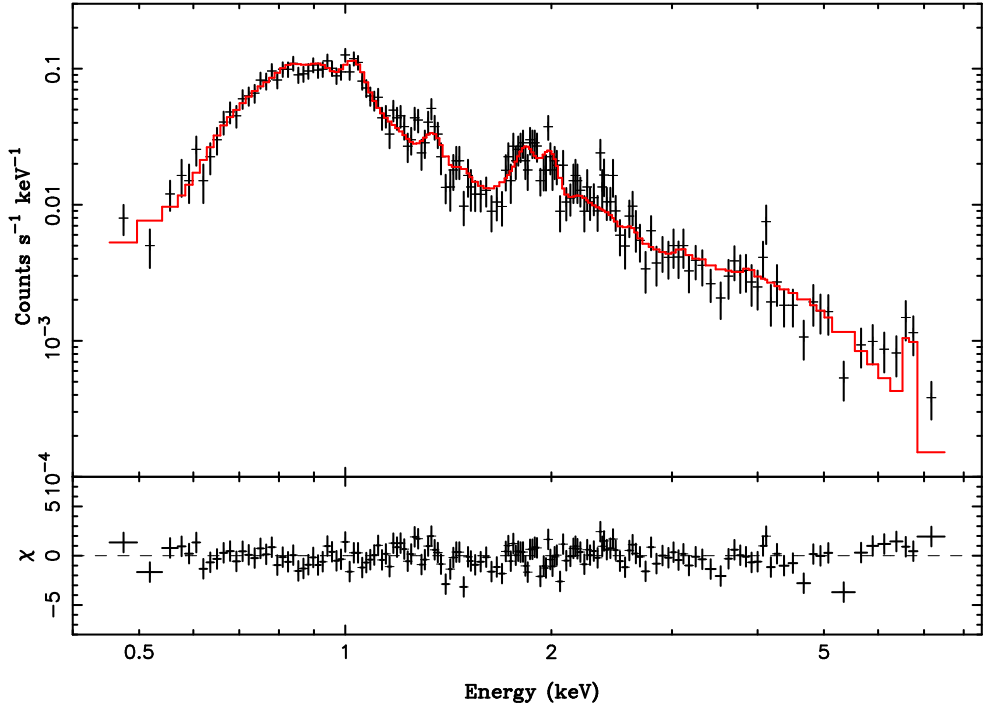}}
\caption{Zero order spectrum of \hd-A (data points) with best fit model
	superimposed (model I of Table~\ref{tab:fit_zoA}; solid line). 
	The best fit model is a two-temperature plasma model with 
	\nh$=1.7 \times	10^{21}$~cm$^{-2}$, $T_1=4.7 \times 10^6$~K 
	(0.4~keV), $T_2=2.7 \times 10^7$~K (2.3~keV).  The X-ray 
	luminosity from this model is $2.9 \times 10^{30}$~erg~s$^{-1}$ 
	(corrected for absorption) in the 0.5-7~keV range.
       \label{fig:zo_spec_A}}
\end{figure*}

As thoroughly discussed in \S\ref{s:DEM_abund}, the high-resolution
spectrum provides additional strong constraints to the fit of the 
X-ray spectrum. Here we use the abundances found from the analysis
of the dispersed spectra and we fit the low-resolutions spectrum
with a 2-T model with abundances fixed to those values. 
This 2-T model, whose parameters are listed in Table~\ref{tab:fit_zoA},
provides a good fit to the integrated zero-order spectrum
of source A, and shown in Figure~\ref{fig:zo_spec_A}.
We also fitted the model with a similar 2-T model but with 
variable abundances (model II in Table~\ref{tab:fit_zoA}), however
the comparison with model I indicates that this model does not
provide a significantly better fit: when abundances are let free 
to vary their best fit values are very similar to the values
found from the analysis of the dispersed spectra which provide 
more robust results because single lines are resolved and used for 
the determination of the element abundances (see \S\ref{s:DEM_abund}). 

Furthermore we explored two more different models: a 2-T model
with fixed abundances as in I but different \nh\ for each 
temperature component (model III), and 3-T model with fixed abundances 
as in model I (model IV).
The former model explores the scenario of two temperature components 
corresponding to different X-ray emitting plasma regions with different
\nh, as likely for example if the two components are emitted by the 
two binary stars. \cite{Skinner96} already investigated this possibility 
when analyzing the ASCA data and found a hot more absorbed component
(T~$\sim 20$~MK, \nh~$\sim 1.7 \times 10^{22}$~cm$^{-3}$) and 
a cool slightly less absorbed component (T~$\sim 2.5$~MK, 
\nh~$\sim 6 \times 10^{21}$~cm$^{-3}$). They tentatively interpreted 
the hot absorbed component as produced by an embedded TTS companion,
and the cool emission as the HAe star emission; the uncertainties 
however were too large to provide any real constraint to this model.
From our data we find that the absorption of the hot component might 
be actually lower than the absorption of the cool component, but 
they are not significantly different within the uncertainties.
The 3-T temperature best fit model has a cool component with parameters
essentially identical to the 2-T models, while the hot component is 
redistributed in two hot components at 20 and 80~MK respectively; the
only parameter slightly different is the lower absorption \nh.
Neither of these two model provides a statistically better fit
to the observed spectrum than the 2-T model with fixed abundances,
therefore we consider model I (Table~\ref{tab:fit_zoA}) the best fit
to the low-resolution spectrum.

Comparing these results with the finding of analysis of previous
X-ray observations of \hd\ we note that the overall characteristics
(T, \lx) of its X-ray emission seem remarkably constant for 
more than a decade, with \lx~$\sim 2 \times 10^{30}$~erg~s$^{-1}$
and at least two temperature components needed to explain its spectrum
(\citealt{Skinner96}, \FHa, \citealt{Skinner04,Stelzer06}).
Both \FHa\ and \cite{Stelzer06} find from the analysis
of \cha-\acis\ data slightly higher temperatures for both T components;
however this can be explained considering that they assume essentially
negligible \nh.
Our model parameters compare well with the findings of \cite{Skinner04}
based on \xmm\ data, the only significant difference being in the 
derived element abundances which \cite{Skinner04} find to be 
significantly above solar for all elements except S, and Fe,
about solar\footnote{They find Ne, Mg, Si, Ca, to be 2.3, 2, 2, 9.9
times solar respectively for their 2-T model, and 2.2, 1.7, 1.6, 6.6
times solar for their 3-T model. As discussed in \S\ref{s:DEM_abund}
the analysis of the high resolution spectra allows to determine 
the element abundances accurately, ruling out such large values.}.

The fits to the low resolution spectra of \hd-A all yield \nh\ values
compatible with the visual extinction derived for this star: the 
\nh\ value spans the range 0.1-0.2$\times 10^{22}$~cm$^{-2}$, which,
using the conversion of \cite{Gorenstein75}, corresponds to $A_V$ 
in the range 0.45-0.9, in good agreement with values found from 
optical studies (e.g., 0.3 of \citealt{vdAncker98}, and 0.8 of 
\citealt{Malfait98}). This is also consistent with previous 
findings of \cite{Skinner04}.

\begin{deluxetable}{lcc}
\tablecolumns{5} 
\tabletypesize{\footnotesize}
\tablecaption{Parameters and uncertainties of models 
	fitting the zero-order spectrum of \hd-A.  
	\label{tab:fit_zoA}}
\tablewidth{0pt}
\tablehead{
 \colhead{Parameter \tablenotemark{a}} & \colhead{I \tablenotemark{b}}  &
 \colhead{II \tablenotemark{c}}   \\
 \colhead{} & \colhead{2T} & \colhead{2T}     
}
\startdata

 \nh$_1$	($10^{22}$~cm$^{-2}$)	    &  0.17 [0.09-0.25]    &  0.12 [0.07-0.19]	\\
 T$_1$		($10^6$~K)		   &  4.7 [4.3-5.4]	  &  5.0 [4.3-5.3]     \\
 EM$_1$		($10^{52}$~cm$^{-3}$)	   &  16.4 [11.6-23.8]    &  11.4 [10.6-12.1]  \\
 \nh$_2$	($10^{22}$~cm$^{-2}$)	    &  =\nh$_1$ 	    &  =\nh$_1$ 	   \\
 T$_2$		($10^6$~K)		   &  27 [24-31]	  &  27 [24-30]        \\
 EM$_2$		($10^{52}$~cm$^{-3}$)	   &  7.1 [6.0-8.5]	  &  6.8 [6.3-7.4]     \\
 T$_3$		($10^6$~K)		   &  - 		  &  -  	       \\
 EM$_3$		($10^{53}$~cm$^{-3}$)	    &  -		   &  -  	       \\
 \fx\ ($10^{-12}$~erg~cm$^{-2}$~s$^{-1}$)   &  1.7 (1.1)	   &  1.5 (1.1)  	\\
 \lx\ ($10^{30}$~erg~s$^{-1}$)		    &  2.9 (1.7)	   &  2.4 (1.7)  	\\
 O/O$_{\odot}$				   & =0.65		  &  0.46 [0.26-0.63]  \\
 Ne/Ne$_{\odot}$			   & =1.1		  &  1.7 [1.4-1.9]     \\
 Mg/Mg$_{\odot}$			   & =0.66		  &  1.1 [0.6-1.3]     \\
 Si/Si$_{\odot}$			   & =0.91		  &  1.2 [0.7-1.5]     \\
 S/S$_{\odot}$				   & =1.24		  &  1.4 [0.6-1.6]     \\
 Fe/Fe$_{\odot}$			   & =0.50		  &  0.42 [0.41-0.53]  \\
 $\chi^2_0$ (d.o.f.)			   & 1.12 (149) 	  &  1.05 (143)       
\enddata 
\tablenotetext{a}{The 90\% confidence intervals are listed in square brackets.
	For X-ray flux and luminosity (\fx, \lx; derived in the 0.5-7.0~keV 
	energy range) we list the value corrected for absorption,
	while the observed (absorbed) values are in parentheses.
	Abundances are relative to solar \citep{Anders89}.}
\tablenotetext{b}{Model with two temperature components and abundances fixed to
	the values found from the analysis of the high resolution spectra (see
	\S\ref{s:DEM_abund}). This is our favored model.}
\tablenotetext{c}{Model with two temperature components and variable abundances.}
\end{deluxetable}

\begin{deluxetable}{lcccc}
\tablecolumns{5} 
\tabletypesize{\footnotesize}
\tablecaption{Parameters and relative uncertainties for isothermal models 
	fitting the zero-order spectra of X-ray sources B, C, D, E.  
	\label{tab:fit_BCDE}}
\tablewidth{0pt}
\tablehead{
 \colhead{Parameter} & \colhead{B} & \colhead{C} & \colhead{D}  &  \colhead{E} 
}
\startdata
 \nh\ ($10^{22}$~cm$^{-2}$)               &  0.0 [0.0-0.8]  &  0.8 [0.-1.7]     &  0.44 [0.26-0.67]  &  3.4 [2.6-5.2] \\
 T ($10^6$~K)                            &  8.4 [1.8-10.]  &  7.0 [2.1-32.]    &  2.8 [2.0-3.6]     &  22. [18-34]   \\
 EM ($10^{53}$~cm$^{-3}$)                & 0.085 [0.064-0.11] &  0.13 [0.012-27.] &  2.9 [0.5-18.]  &  1.8 [0.5-3.8] \\
 \fx\ ($10^{-13}$~erg~cm$^{-2}$~s$^{-1}$) &  0.35 (0.35)    &  0.4 (0.09)       &  6.5 (0.73)        &  7.1 (1.9)     \\
 \lx\ ($10^{30}$~erg~s$^{-1}$)            &  0.057 (0.057)  &  0.07 (0.0015)    &  1.0 (0.12)        &  1.1 (0.31)    \\
 $\chi^2_0$                              &  1.45           &  1.24             &  1.54              &  0.86          \\
 d.o.f.                                  &  13            &  2                 &  6                 &  20            \\
 \enddata 
\tablecomments{The 90\% confidence intervals are listed in square brackets.
	For X-ray flux and luminosity (\fx, \lx; derived in the 0.5-7.0~keV 
	energy range) we list the value corrected for absorption,
	while the observed (absorbed) values are in parentheses.
	For all sources the global metal abundances are fixed to 0.3 times
	solar.}
\end{deluxetable}

\begin{figure*}[!tbhp]
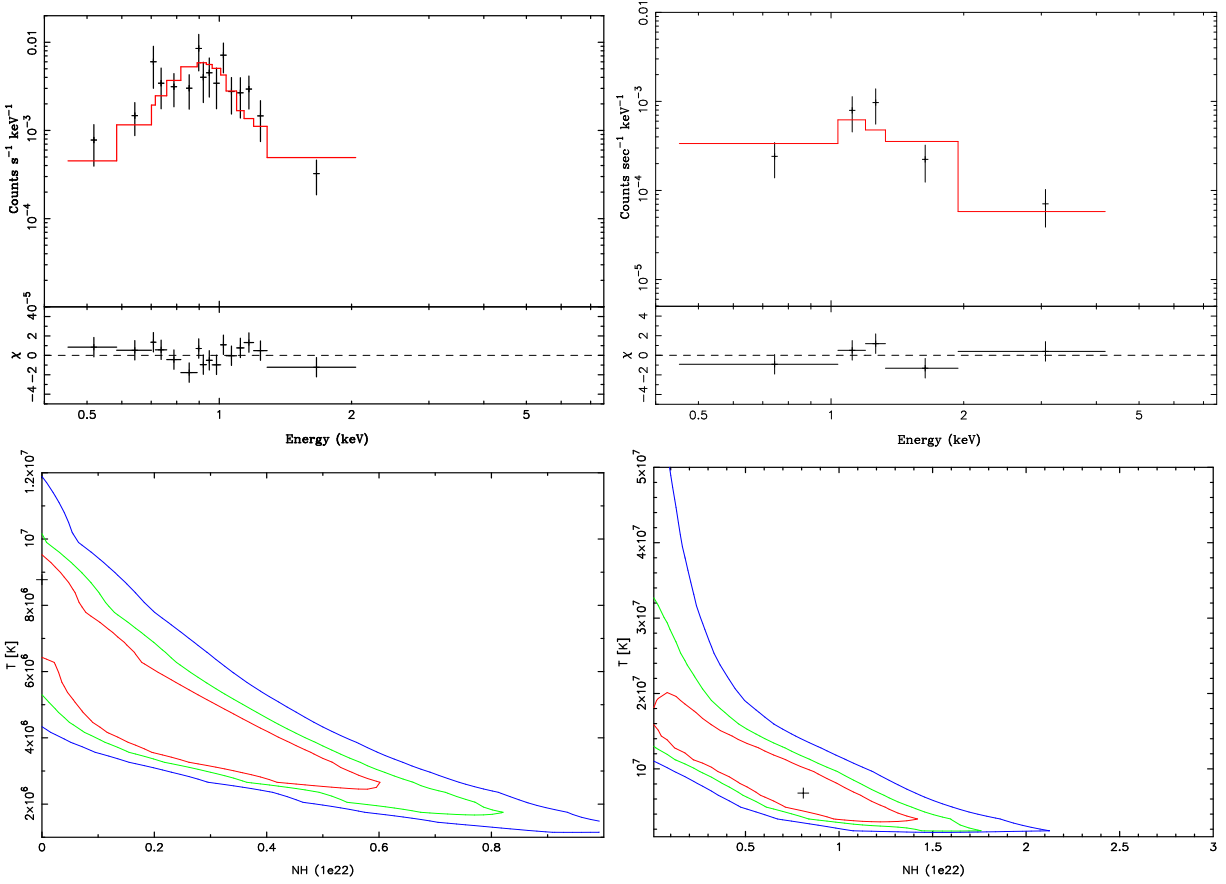

\centerline{\psfig{figure=f6a.ps,width=8cm,angle=-90}
	\psfig{figure=f6b.ps,width=8cm,angle=-90}}
\centerline{\psfig{figure=f6c.ps,width=8cm,height=5.7cm,angle=-90}
	\psfig{figure=f6d.ps,width=8cm,height=5.7cm,angle=-90}}
\caption{{\em Top:} Zero-order spectrum of \hd-B ({\em left}) and C({\em right}), 
	rebinned to a minimum 5 counts per bin, with corresponding
	best fit model.  The parameters of the best fit models are 
	listed in Table~\ref{tab:fit_BCDE}.
       {\em Bottom:} Confidence contours in the \nh-T space
       (68\%, 90\%, 98\%) for source B ({\em left}) and C ({\em right}).
       \label{fig:zo_spec_BC}}
\end{figure*}

\begin{figure*}[!tbhp]
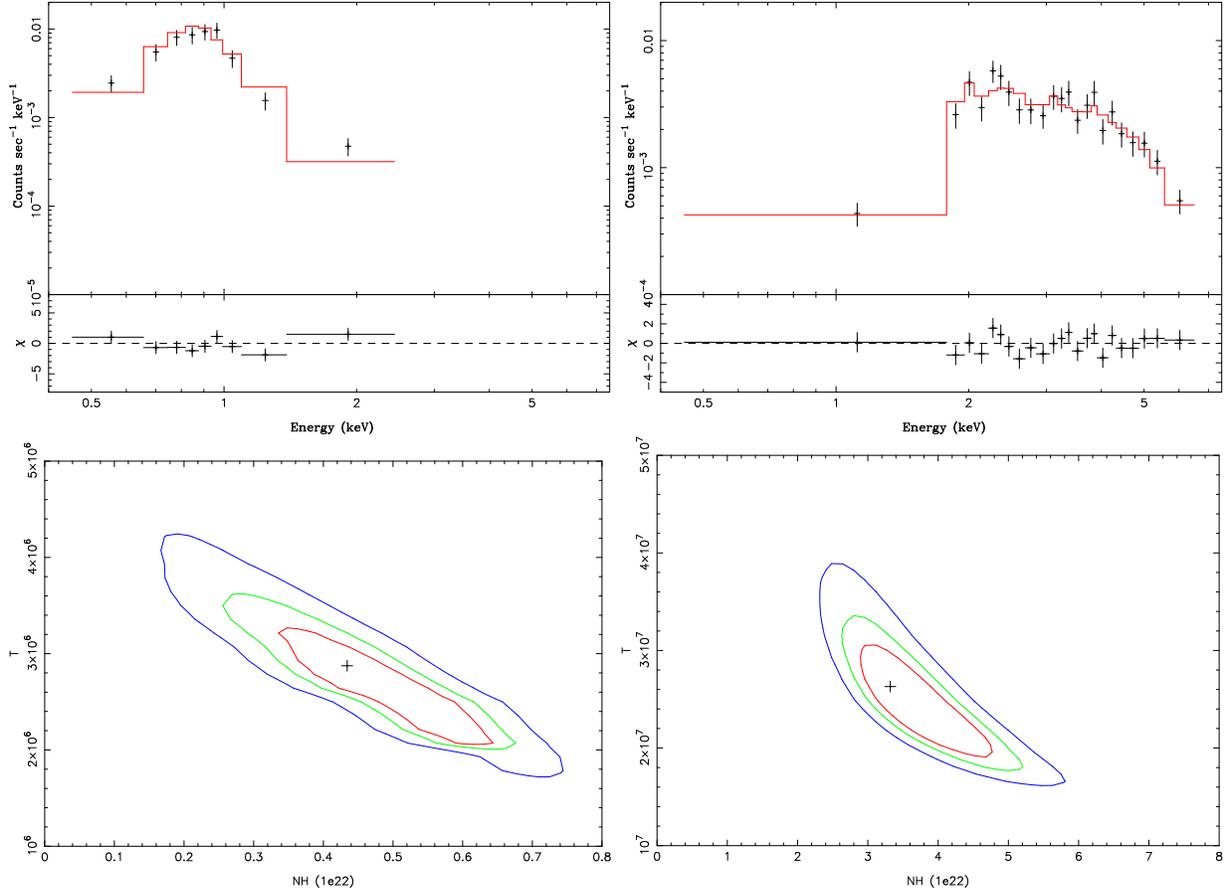

\centerline{\psfig{figure=f7a.ps,width=8cm,angle=-90}
	   \psfig{figure=f7b.ps,width=8cm,angle=-90}}
\centerline{\psfig{figure=f7c.ps,width=8cm,angle=-90}
	   \psfig{figure=f7d.ps,width=8cm,angle=-90}}
\caption{{\em Top:} Zero order spectrum of \hd-D ({\em left}) 
       and E ({\em right}) with best fit model.  The parameters
       of the best fit models are listed in Table~\ref{tab:fit_BCDE}.
       {\em Bottom:} Confidence contours in the \nh-T space
       (68\%, 90\%, 98\%) for D ({\em left}) and E ({\em right}).
       \label{fig:zo_spec_DE}}
\end{figure*}

The zero-order spectra of the other four sources are shown in 
Figure~\ref{fig:zo_spec_BC}~and~\ref{fig:zo_spec_DE}, and the 
parameters of the fits with corresponding uncertainties are 
listed in Table~\ref{tab:fit_BCDE}.

The left panel of Figure~\ref{fig:zo_spec_BC} shows the zero-order
spectrum of source B with the corresponding best fit model. We
find that a model with a single temperature at $\sim 1$~keV and 
zero absorption reproduces well enough the observed spectrum.
The parameters of the X-ray spectrum are rather typical of similar
M-type TTS (see discussion in \S\ref{s:discuss}).
However the relatively low number of counts ($\sim 150$) obtained for 
this source does not allow to constrain the parameters very well, 
as demonstrated in the confidence contours in the \nh-T 
parameter space shown in the panel below the spectrum plot in 
Figure~\ref{fig:zo_spec_BC}.
     
Source C is a very weak source for which very scarce information is 
available. On the basis of the X-ray observation and the (lack
of) optical data, \FHa\ proposed a possible identification as a 
brown dwarf; the association with the young stellar group of \hd\ 
is uncertain. In our 145~ks observation we gathered about 54 net counts
for source C (subtracting a background estimated on a nearby extraction
region); the spectrum and the best fit model are shown in 
Figure~\ref{fig:zo_spec_BC}. The poor statistics do not provide 
stringent constraints on the spectral properties of the X-ray emission
of this source. Within the uncertainties, we find parameters 
compatible with the emission by a brown dwarf at the distance of \hd\ 
confirming the findings of \FHa.

For source D, and E, which are identified as T Tauri stars associated 
to \hd, we have obtained about 250 and 540 zero-order counts 
respectively, over the 145~ks observation. 
As already apparent in the color coded image (Fig.~\ref{fig:img}), 
source E is heavily absorbed and only the hard photons are detected. 
The spectral fit indicates in fact a large \nh\ of 
$3.4 \times 10^{22}$~cm$^{-2}$, much larger than the interstellar 
absorption and that can be therefore ascribed to local absorption from 
circumstellar material. This source is identified as a K3 classical 
T Tauri star, and the infrared excess detected by G04 provide further 
evidence of presence of large amounts of circumstellar material. 
Source D is a lower mass T Tauri star, of spectral-type M, border-line
classical T Tauri star as indicated by the $H \alpha$ equivalent width
very similar to source E (G04). At variance with respect to source E,
infrared measurement do not indicate large infrared excess (G04;
however the measurements are characterized by poorer statistics for
this source and therefore the limits are not very stringent).
Its X-ray spectrum is very soft for its emission level (its
\lx$=10^{30}$~erg~s$^{-1}$ is rather large for M-type TTS;
see \S\ref{s:discuss} for a more detailed discussion),
and it is fit well with an isothermal model with $T= 3 \times 10^6$~K 
(0.24~keV).
These spectral parameters for source D, and E, are well determined
within the statistical uncertainties as clear from the confidence 
contours in the \nh-T space shown in Figure~\ref{fig:zo_spec_DE}
(bottom panels).

\subsection{\hetgs\ spectrum of \hd-A}
\label{s:hetgspec}

A goal of our observing program was to obtain a high
resolution X-ray spectrum of the Herbig Ae star with high 
signal-to-noise ratio, in order to investigate the X-ray 
production mechanisms in intermediate mass pre-main sequence 
stars.
The HETGS spectrum of sources A and B cannot be resolved spatially 
in the dispersed spectra. However, our zero-order analysis 
(\S\ref{s:lowres}) indicates that component B's contribution is 
$\lesssim 10$\% that of A.   Henceforth we will refer to the 
spectrum as if from A alone, but it should be kept in mind that 
there is some level of contamination from B.  Other sources in 
the field are distant enough that order-sorting results in unique 
determination in \meg-\heg\ crossed-order regions.
In the 145~ks \hetgs\ observation we obtained slightly more than
5000 total counts in the \heg\ and \meg\ -1,+1 orders.

\begin{figure*}[!tbhp]
\centerline{\psfig{figure=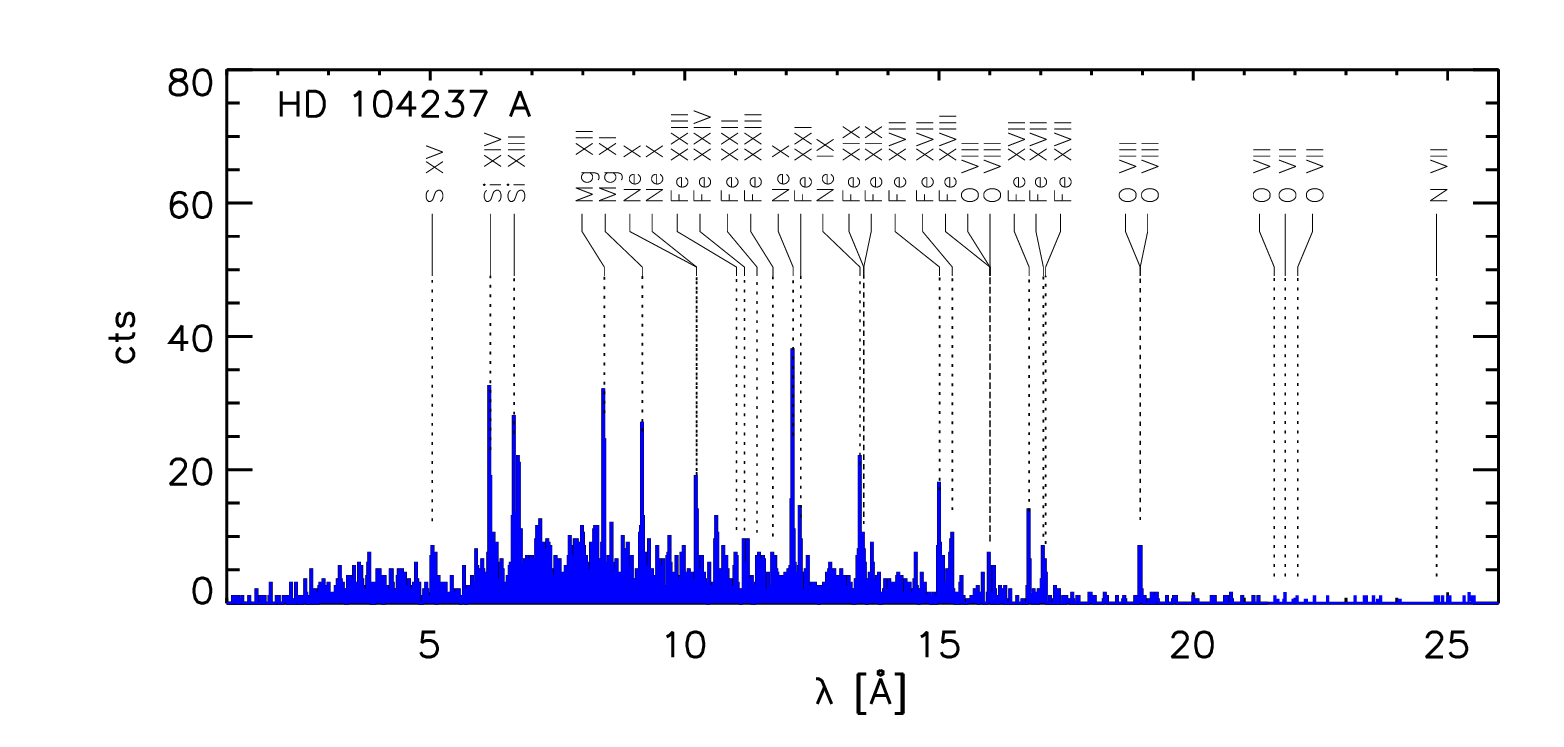,width=17cm}\vspace{-0.5cm}}
\caption{\hd-A \cha-\hetgs\ spectrum (\heg\ -1,+1, and \meg\ -1,+1 
	summed; binsize=0.02\AA) integrated over the
	145~ks observation.
       The most prominent lines are labeled. \label{fig:hdA}}
\end{figure*}

In Figure~\ref{fig:hdA} we show the high resolution spectrum 
of \hd-A, obtained by summing \heg\ and \meg\ spectra (-1 and 
+1 orders).
The spectrum is dominated by line emission of H-like and He-like
ions of Si, Mg, Ne, by \fexvii\ transitions around 15\AA\ and 17\AA, 
and by transitions of H-like oxygen. Also, some continuum emission 
is present as well as transitions of highly ionized Fe around 
$\sim 11$~\AA\ both indicating the presence of significant
amounts of hot ($T \gtrsim 2 \times 10^7$~K) plasma.

The measured line fluxes together with the statistical errors, are 
listed in Table~\ref{tab:linefluxes}.
 
These fluxes were determined by fitting each narrow ($\sim 0.5$ \AA)
region of interest with a sum of Gaussians and a plasma model
continuum, all multiplied by a function of absorption by neutral
gas with solar abundances.
The model was folded through the instrument response for each of the
negative and positive first orders for each grating, then the
positive and negative orders were combined before computing the
statistic.  \heg\ and \meg\ orders were combined when feasible
(roughly between 3-16 \AA, where the spectral coverage overlaps with
significant sensitivity).  The line widths were usually frozen at a
negligible intrinsic value (in \S\ref{s:line_prof} we describe an
independent analysis of line profiles).  Only for \eli{O}{8} 19 \AA\
did we need to free the width parameter; this is because of the high
dispersion at this wavelength and the fact that the
H-Ly$\alpha$-like lines are doublets, with a separation of $0.006$
\AA\ for \eli{O}{8}.

The continuum model was determined by a prior fit of a three
temperature-component continuum model to relatively line-free
regions of the spectrum, using the Astrophysical Plasma Emission
Database continuum emissivities (APED\footnote{APED is available
  from http://cxc.harvard.edu/atomdb/}; \citealt{Smith01}).  This
continuum then became an {\it a priori} determined component for the
line fits.  For the absorption, we used the {\tt wabs} function from
the {\tt XSPEC} module in ISIS and a fixed value of \nh\ (see
\S\ref{s:hetgspec}).  From these fits we obtained a single
(unabsorbed) flux value as well as a wavelength.

We did not physically merge any data or response files into new
files.  Combination of the $\pm1^\mathrm{st}$ orders was done
dynamically.  There are many advantages to this technique.  Foremost
is that the individual responses are still applied to the model to
compute the predicted counts for each order.  Second, individual
orders can be noticed and ignored as required for each region.
Third, the combination after folding, and before the statistic,
increases the signal-to-noise per bin (which does not occur for a
joint fit).  Fourth, the method is transparent to combination of
orders from one observation or from multiple observations, since all
responses, counts spectra, and exposure times are still unique.  The
visualization in Figure~\ref{fig:hdA} shows the summed, combined,
and rebinned \heg\ plus \meg\ counts spectrum, but this was not the
object directly fit.

Since we have a fairly low-counts spectrum, we also regridded the
data (and corresponding response) to the \heg\ binsize ($0.0025$
\AA).  This allowed us to combine \heg\ and \meg\ spectra, but to
keep \heg\ resolution in the \heg\ counts spectrum where feasible.
Otherwise, we grouped data by an integral numbers of bins
independently for each feature.

For \fexxv\ ($1.85$ \AA), we also included the zeroth order spectrum
in the fit, since the zeroth order has a comparable number of
counts, and though lower resolution, there are no significant
confusing lines.

These dynamic functions for combining data, regridding responses and
data, and for grouping data are standard features of ISIS.

\begin{deluxetable}{lccrrrrrr}
 \tabletypesize{\scriptsize}
 \tablecaption{Line Measurements \label{tab:linefluxes} }
 \tablewidth{0pt}
 \tablehead{
   \colhead{Ion} &
   \colhead{Use\tablenotemark{a}} & 
   \colhead{$\overline{\log T}$\tablenotemark{b}}&
   \colhead{$\lambda_t$\tablenotemark{c}}&
   \colhead{$\lambda_o$\tablenotemark{d}}&
   \colhead{$f_l$\tablenotemark{e}}&
   \colhead{$f_t$\tablenotemark{f}}&
   \colhead{$\delta f$\tablenotemark{g}}&
   \colhead{$\delta\chi$\tablenotemark{h}}
 }
 \startdata
\fexxv\    & E & 7.8 & 1.861  & 1.868  (4.8) & 1.15 (0.34)  & 0.712  & 0.441  & 1.30 \\
\arxviii\  & E & 7.7 & 3.734  & 3.738  (15.) & 0.18 (0.23)  & 0.154  & 0.025  & 0.11 \\
\arxvii\   & E & 7.4 & 3.949  & 3.949  (8.5) & 0.56 (0.47)  & 0.223  & 0.335  & 0.71 \\
\arxvii\   &   & 7.3 & 3.968  & 3.953  (15.) & 0.001 (0.37) & 0.061  & -0.060 & -0.16 \\
\sxvi\     &   & 7.5 & 3.992  & 4.007  (4.7) & 0.26 (0.27)  & 0.249  & 0.006  & 0.02 \\
\sxv\      & E & 7.3 & 4.088  & 4.088  (15.) & 0.07 (0.18)  & 0.064  & 0.006  & 0.03 \\
\sxvi\     & E & 7.6 & 4.730  & 4.729  (11.) & 1.65 (0.85)  & 1.221  & 0.431  & 0.51 \\
\sixiv\    & E & 7.4 & 4.947  & 4.947  (-) & 0.11 (0.19)  & 0.086  & 0.019  & 0.10 \\
\sxv\      & E & 7.2 & 5.039  & 5.040  (2.0) & 2.22 (0.61)  & 1.670  & 0.551  & 0.91 \\
\sxv\      & E & 7.2 & 5.065  & 5.065  (-) & 0.23 (0.32)  & 0.390  & -0.160 & -0.49 \\
\sxv\      & E & 7.2 & 5.102  & 5.107  (5.8) & 1.03 (0.46)  & 0.576  & 0.453  & 0.99 \\
\sixiv\    & E & 7.4 & 5.217  & 5.219  (2.8) & 0.86 (0.46)  & 0.407  & 0.450  & 0.99 \\
\sixiii\   &   & 7.1 & 5.285  & 5.285  (-) & 0.002 (0.18) & 0.062  & -0.059 & -0.33 \\
\sixiii\   & E & 7.1 & 5.405  & 5.405  (-) & 0.14 (0.29)  & 0.136  & 0.000  & 0.00 \\
\sixiv\    & E & 7.4 & 6.183  & 6.181  (2.1) & 3.51 (0.88)  & 3.053  & 0.462  & 0.53 \\
\sixiv\    &   & 7.1 & 6.265  & 6.256  (15.) & 0.45 (0.57)  & 0.072  & 0.374  & 0.65 \\
\mgxii\    &   & 7.2 & 6.580  & 6.580  (-) & 0.001 (0.11) & 0.063  & -0.062 & -0.59 \\
\sixiii\   & E & 7.0 & 6.648  & 6.650  (1.1) & 3.63 (0.48)  & 3.609  & 0.018  & 0.04 \\
\sixiii\   &   & 7.0 & 6.687  & 6.687  (-) & 0.18 (0.21)  & 0.688  & -0.506 & -2.44 \\
\sixiii\   &   & 6.8 & 6.720  & 6.720  (-) & 0.60 (0.28)  & 0.035  & 0.561  & 2.02 \\
\sixiii\   & E & 7.0 & 6.740  & 6.743  (1.3) & 2.06 (0.35)  & 1.465  & 0.594  & 1.69 \\
\mgxii\    & E & 7.2 & 7.106  & 7.107  (5.5) & 0.83 (0.32)  & 0.458  & 0.370  & 1.15 \\
\alxiii\   &   & 7.4 & 7.171  & 7.167  (11.) & 0.40 (0.28)  & 0.686  & -0.284 & -1.02 \\
\mgxi\     &   & 6.9 & 7.310  & 7.317  (6.8) & 0.56 (0.31)  & 0.051  & 0.508  & 1.65 \\
\fexxii\   & E & 7.1 & 7.681  & 7.667  (15.) & 0.06 (0.19)  & 0.037  & 0.020  & 0.11 \\
\alxii\    & E & 7.0 & 7.757  & 7.761  (15.) & 0.62 (0.37)  & 0.243  & 0.374  & 1.00 \\
\alxii\    &   & 6.9 & 7.805  & 7.790  (15.) & 0.06 (0.19)  & 0.079  & -0.015 & -0.08 \\
\mgxi\     & E & 6.9 & 7.850  & 7.852  (5.8) & 0.57 (0.33)  & 0.349  & 0.218  & 0.66 \\
\alxii\    & E & 6.9 & 7.872  & 7.879  (15.) & 0.26 (0.30)  & 0.220  & 0.038  & 0.12 \\
\fexxiii\  &   & 7.2 & 7.901  & 7.897  (15.) & 0.33 (0.35)  & 0.062  & 0.267  & 0.76 \\
\fexxiv\   & E & 7.4 & 7.986  & 7.988  (10.) & 0.35 (0.38)  & 0.192  & 0.161  & 0.43 \\
\fexxiv\   & E & 7.4 & 7.996  & 8.011  (7.5) & 0.28 (0.29)  & 0.097  & 0.179  & 0.63 \\
\fexxiii\  & E & 7.2 & 8.304  & 8.300  (11.) & 0.28 (0.33)  & 0.229  & 0.052  & 0.16 \\
\fexxiv\   & E & 7.4 & 8.316  & 8.317  (15.) & 0.20 (0.31)  & 0.217  & -0.018 & -0.06 \\
\fexxiv\   &   & 7.4 & 8.376  & 8.389  (-) & 0.34 (0.34)  & 0.083  & 0.253  & 0.75 \\
\mgxii\    & E & 7.2 & 8.422  & 8.422  (1.6) & 3.65 (0.50)  & 3.626  & 0.026  & 0.05 \\
\fexxi\    & E & 7.1 & 8.574  & 8.567  (4.9) & 0.58 (0.21)  & 0.188  & 0.395  & 1.88 \\
\fexxiii\  & E & 7.2 & 8.815  & 8.816  (3.9) & 0.69 (0.28)  & 0.247  & 0.447  & 1.59 \\
\fexxii\   & E & 7.1 & 8.975  & 8.976  (6.6) & 0.42 (0.28)  & 0.321  & 0.095  & 0.34 \\
\mgxi\     & E & 6.8 & 9.169  & 9.169  (1.3) & 4.24 (0.78)  & 3.046  & 1.195  & 1.54 \\
\fexxi\    &   & 7.1 & 9.194  & 9.194  (-) & 1.14 (0.44)  & 0.144  & 0.994  & 2.24 \\
\mgxi\     &   & 6.8 & 9.230  & 9.230  (-) & 1.39 (0.50)  & 0.489  & 0.895  & 1.80 \\
\mgxi\     &   & 6.8 & 9.314  & 9.315  (2.7) & 2.20 (0.65)  & 1.447  & 0.755  & 1.17 \\
\fexxii\   &   & 7.1 & 9.393  & 9.393  (-) & 0.001 (0.20) & 0.072  & -0.071 & -0.36 \\
\nex\      &   & 7.0 & 9.481  & 9.472  (3.2) & 0.48 (0.25)  & 0.379  & 0.096  & 0.38 \\
\fexix\    &   & 6.9 & 9.695  & 9.680  (3.8) & 0.72 (0.36)  & 0.122  & 0.601  & 1.66 \\
\nex\      & E & 7.0 & 9.708  & 9.708  (-) & 1.35 (0.61)  & 0.858  & 0.492  & 0.81 \\
\fexx\     & E & 7.0 & 9.727  & 9.727  (-) & 0.51 (0.39)  & 0.143  & 0.362  & 0.94 \\
\nixix\    & E & 6.8 & 10.110 & 10.111 (9.6) & 0.31 (0.36)  & 0.233  & 0.074  & 0.20 \\
\fexx\     & E & 7.0 & 10.120 & 10.127 (14.) & 0.39 (0.27)  & 0.161  & 0.232  & 0.87 \\
\nex\      & E & 7.0 & 10.239 & 10.240 (1.7) & 3.98 (0.83)  & 2.817  & 1.165  & 1.41 \\
\fexxiv\   & E & 7.4 & 10.619 & 10.619 (-) & 1.36 (0.60)  & 1.576  & -0.217 & -0.36 \\
\fexix\    & E & 6.9 & 10.641 & 10.641 (-) & 1.05 (0.78)  & 0.273  & 0.772  & 1.00 \\
\fexix\    & E & 6.9 & 10.649 & 10.649 (-) & 0.42 (0.60)  & 0.264  & 0.152  & 0.25 \\
\fexxiv\   & E & 7.4 & 10.663 & 10.663 (-) & 0.80 (0.58)  & 0.830  & -0.026 & -0.04 \\
\fexxiii\  &   & 7.2 & 10.981 & 10.983 (13.) & 0.39 (0.47)  & 1.443  & -1.053 & -2.26 \\
\neix\     & E & 6.7 & 11.001 & 11.001 (-) & 1.23 (0.61)  & 0.418  & 0.812  & 1.33 \\
\fexxiii\  & E & 7.2 & 11.019 & 11.019 (-) & 1.34 (0.72)  & 0.953  & 0.382  & 0.53 \\
\fexxiv\   & E & 7.4 & 11.029 & 11.029 (-) & 1.25 (0.65)  & 1.047  & 0.205  & 0.32 \\
\fexvii\   &   & 6.7 & 11.131 & 11.123 (15.) & 0.001 (0.09) & 0.810  & -0.809 & -8.82 \\
\fexxiv\   & E & 7.4 & 11.176 & 11.176 (2.3) & 2.72 (0.77)  & 1.915  & 0.799  & 1.04 \\
\fexvii\   & E & 6.7 & 11.254 & 11.251 (2.8) & 2.54 (0.76)  & 1.146  & 1.397  & 1.84 \\
\fexviii\  & E & 6.8 & 11.326 & 11.321 (4.5) & 1.61 (0.64)  & 0.838  & 0.766  & 1.19 \\
\fexviii\  & E & 6.8 & 11.527 & 11.524 (6.8) & 0.99 (0.61)  & 0.890  & 0.101  & 0.17 \\
\neix\     & E & 6.6 & 11.544 & 11.548 (4.9) & 1.67 (0.69)  & 1.362  & 0.304  & 0.44 \\
\fexxiii\  &   & 7.2 & 11.736 & 11.741 (1.8) & 4.96 (0.99)  & 3.323  & 1.637  & 1.66 \\
\fexxii\   & E & 7.1 & 11.770 & 11.773 (2.3) & 3.34 (1.01)  & 3.561  & -0.223 & -0.22 \\
\nex\      & E & 6.9 & 12.135 & 12.131 (1.0) & 24.0 (2.9)   & 24.246 & -0.254 & -0.09 \\
\fexxiii\  &   & 7.2 & 12.161 & 12.150 (4.2) & 4.88 (2.05)  & 1.934  & 2.943  & 1.43 \\
\fexvii\   & E & 6.7 & 12.266 & 12.265 (3.4) & 3.60 (1.37)  & 3.113  & 0.483  & 0.35 \\
\fexxi\    & E & 7.1 & 12.284 & 12.284 (2.1) & 7.82 (1.75)  & 7.538  & 0.282  & 0.16 \\
\fexx\     & E & 7.0 & 13.385 & 13.396 (6.5) & 1.53 (1.17)  & 1.043  & 0.487  & 0.42 \\
\fexix\    & E & 6.9 & 13.423 & 13.423 (-) & 2.57 (1.38)  & 0.559  & 2.008  & 1.46 \\
\neix\     & E & 6.6 & 13.447 & 13.449 (1.5) & 15.2 (2.7)   & 12.577 & 2.657  & 1.00 \\
\fexix\    & E & 6.9 & 13.462 & 13.462 (-) & 2.34 (1.73)  & 1.274  & 1.066  & 0.62 \\
\fexix\    &   & 6.9 & 13.497 & 13.497 (-) & 0.36 (0.74)  & 2.245  & -1.884 & -2.54 \\
\fexix\    & E & 6.9 & 13.518 & 13.519 (1.9) & 9.60 (2.34)  & 4.958  & 4.646  & 1.98 \\
\neix\     &   & 6.6 & 13.552 & 13.554 (3.3) & 3.95 (1.57)  & 1.904  & 2.041  & 1.30 \\
\fexix\    & E & 6.9 & 13.645 & 13.653 (-) & 1.49 (2.46)  & 0.801  & 0.687  & 0.28 \\
\neix\     & E & 6.6 & 13.699 & 13.694 (6.4) & 6.91 (3.20)  & 6.299  & 0.607  & 0.19 \\
\fexx\     & E & 7.0 & 13.767 & 13.762 (10.) & 3.77 (3.37)  & 0.917  & 2.856  & 0.85 \\
\fexix\    & E & 6.9 & 13.795 & 13.790 (7.6) & 3.90 (3.33)  & 2.036  & 1.863  & 0.56 \\
\fexvii\   & E & 6.7 & 13.825 & 13.827 (11.) & 3.92 (3.26)  & 2.787  & 1.129  & 0.35 \\
\fexviii\  & E & 6.8 & 14.208 & 14.207 (8.3) & 9.60 (4.83)  & 12.652 & -3.051 & -0.63 \\
\fexviii\  & E & 6.8 & 14.256 & 14.249 (13.) & 3.57 (3.68)  & 2.445  & 1.122  & 0.31 \\
\fexx\     & E & 7.0 & 14.267 & 14.268 (13.) & 3.59 (4.22)  & 1.561  & 2.030  & 0.48 \\
\fexviii\  & E & 6.8 & 14.343 & 14.355 (15.) & 1.52 (3.24)  & 1.429  & 0.090  & 0.03 \\
\fexviii\  & E & 6.8 & 14.373 & 14.374 (5.4) & 9.40 (3.77)  & 3.077  & 6.324  & 1.68 \\
\fexviii\  &   & 6.8 & 14.425 & 14.410 (15.) & 3.49 (2.96)  & 0.668  & 2.821  & 0.95 \\
\fexviii\  &   & 6.8 & 14.534 & 14.547 (6.1) & 4.36 (3.49)  & 2.365  & 1.999  & 0.57 \\
\fexix\    & E & 6.9 & 14.664 & 14.667 (11.) & 4.95 (3.14)  & 1.558  & 3.389  & 1.08 \\
\oviii\    & E & 6.7 & 14.821 & 14.817 (4.4) & 4.37 (1.78)  & 1.174  & 3.193  & 1.79 \\
\fexvii\   & E & 6.7 & 15.014 & 15.014 (3.4) & 32.8 (7.4)   & 39.169 & -6.377 & -0.86 \\
\fexix\    & E & 6.9 & 15.079 & 15.087 (9.2) & 6.55 (6.59)  & 1.809  & 4.738  & 0.72 \\
\oviii\    & E & 6.7 & 15.176 & 15.178 (15.) & 3.26 (6.12)  & 2.741  & 0.515  & 0.08 \\
\fexix\    & E & 6.9 & 15.198 & 15.209 (11.) & 7.39 (6.26)  & 1.514  & 5.875  & 0.94 \\
\fexvii\   & E & 6.7 & 15.261 & 15.261 (5.5) & 17.3 (6.8)   & 11.351 & 5.939  & 0.87 \\
\fexvii\   & E & 6.7 & 15.453 & 15.451 (15.) & 4.26 (4.53)  & 1.484  & 2.778  & 0.61 \\
\fexviii\  &   & 6.8 & 15.494 & 15.488 (15.) & 2.17 (3.32)  & 0.386  & 1.784  & 0.54 \\
\fexviii\  & E & 6.8 & 15.625 & 15.625 (15.) & 2.51 (3.69)  & 3.604  & -1.096 & -0.30 \\
\fexviii\  &   & 6.8 & 15.759 & 15.763 (15.) & 3.37 (4.31)  & 0.497  & 2.876  & 0.67 \\
\fexviii\  & E & 6.8 & 15.824 & 15.826 (15.) & 4.60 (5.04)  & 2.251  & 2.346  & 0.47 \\
\fexviii\  & E & 6.8 & 15.870 & 15.875 (15.) & 5.22 (5.49)  & 1.200  & 4.018  & 0.73 \\
\oviii\    &   & 6.7 & 16.006 & 16.004 (2.2) & 21.7 (5.0)   & 9.436  & 12.242 & 2.43 \\
\fexviii\  & E & 6.8 & 16.071 & 16.075 (6.1) & 4.22 (2.39)  & 5.304  & -1.087 & -0.45 \\
\fexix\    &   & 6.9 & 16.110 & 16.095 (2.5) & 7.60 (2.97)  & 2.642  & 4.963  & 1.67 \\
\fexviii\  &   & 6.8 & 16.159 & 16.150 (4.0) & 6.05 (2.49)  & 2.132  & 3.922  & 1.57 \\
\fexvii\   & E & 6.7 & 16.780 & 16.779 (4.0) & 41.4 (11.6)  & 21.558 & 19.810 & 1.71 \\
\fexvii\   & E & 6.7 & 17.051 & 17.052 (6.3) & 29.4 (10.9)  & 26.525 & 2.838  & 0.26 \\
\fexvii\   & E & 6.7 & 17.096 & 17.100 (4.4) & 37.8 (11.2)  & 24.814 & 13.014 & 1.16 \\
\fexviii\  &   & 6.8 & 17.623 & 17.635 (4.7) & 6.14 (3.39)  & 4.127  & 2.010  & 0.59 \\
\ovii\     &   & 6.4 & 17.768 & 17.762 (8.1) & 4.87 (3.26)  & 0.663  & 4.203  & 1.29 \\
\fexvii\   & E & 6.6 & 17.793 & 17.778 (15.) & 0.001 (1.48) & 0.000  & 0.000  & 0.00 \\
\ovii\     &   & 6.4 & 18.627 & 18.627 (-) & 0.001 (1.24) & 2.329  & -2.328 & -1.87 \\
\caxviii\  & E & 7.1 & 18.691 & 18.691 (-) & 0.46 (3.05)  & 0.644  & -0.185 & -0.06 \\
\oviii\    & E & 6.7 & 18.970 & 18.971 (2.2) & 104. (17.)   & 89.094 & 14.826 & 0.85 \\
\caxvi\    &   & 6.7 & 20.859 & 20.869 (12.) & 7.09 (7.70)  & 0.043  & 7.046  & 0.92 \\
\nvii\     &   & 6.5 & 20.910 & 20.881 (30.) & 0.001 (5.28) & 1.958  & -1.958 & -0.37 \\
\ovii\     & E & 6.3 & 21.601 & 21.605 (10.) & 20.0 (16.3)  & 28.511 & -8.519 & -0.52 \\
\ovii\     &   & 6.3 & 21.802 & 21.817 (6.4) & 17.3 (15.3)  & 3.861  & 13.392 & 0.88 \\
\ovii\     &   & 6.3 & 22.098 & 22.083 (15.) & 2.20 (9.83)  & 16.800 & -14.603 & -1.49 \\
\nvii\     & E & 6.5 & 24.782 & 24.797 (15.) & 20.4 (14.3)  & 13.131 & 7.253  & 0.51 \\
\tableline
\enddata
\tablenotetext{a}{The label E indicates the lines used for the emission measure reconstruction.}
\tablenotetext{b}{Average logarithmic temperature (kelvins) of formation.}
\tablenotetext{c}{Theoretical wavelengths of identification (from APED), 
			in \AA\ units. If the line is a multiplet, we list 
			the wavelength of the strongest component.}
\tablenotetext{d}{Measured wavelength, in angstrom units (uncertainty is in m\AA).
			Errors equal to (-) indicate that the line wavelength was frozen
			because the feature is very weak. }
\tablenotetext{e}{Measured line fluxes in $10^{-6}$~photons~cm$^{-2}$~s$^{-1}$ 
		       for \heg+\meg\ first orders, with 1$\sigma$ uncertainties, assuming
		       \nh$=1.5 \times 10^{21}$~cm$^{-2}$.}
\tablenotetext{f}{Fluxes (in $10^{-6}$~photons~cm$^{-2}$~s$^{-1}$) predicted using 
		       the emission measure distribution model (see Fig.~\ref{fig:dem}).}
\tablenotetext{g}{Line flux residuals, $\delta f = f_l - f_t$.}
\tablenotetext{h}{Discrepancy between observed and predicted fluxes 
		       in units of $\sigma$.}
\end{deluxetable}

\subsubsection{Emission Measure Distribution and Abundances}
\label{s:DEM_abund}

\begin{figure*}[!tbhp]
\centerline{\psfig{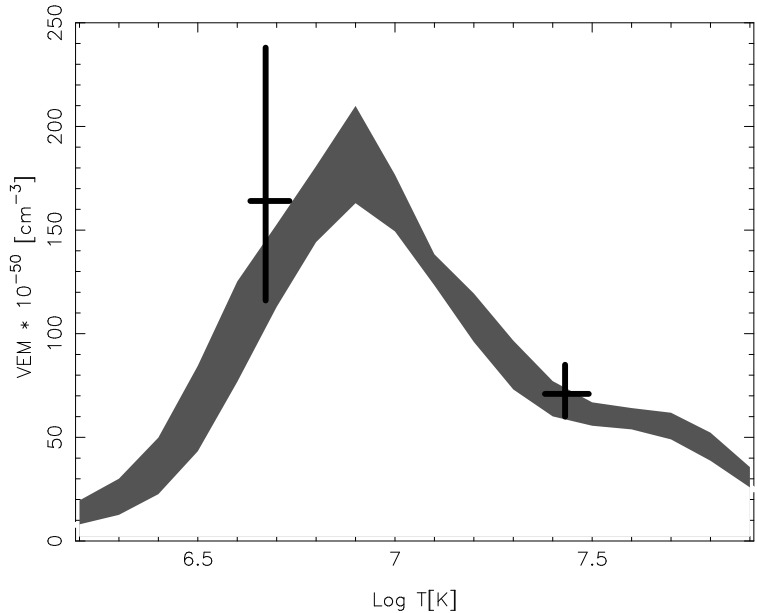}}
\caption{Emission measure distribution inferred from the measured line 
	fluxes.   
	Crosses show the EM component derived from the $2T$ fit to 
	zero-order spectra. Since we are plotting the emission measure 
	integrated over the temperature bins, we applied an arbitrary 
	scaling factor of 0.1 to the 2T models in order to compare 
	the relative weights with the shape of the EMD; total volume 
	emission measures for the two models, $2T$ and $EMD$, are 
	comparable, as they should be.
       \label{fig:dem}}
\end{figure*}

\begin{figure*}[!tbhp]
\centerline{\psfig{figure=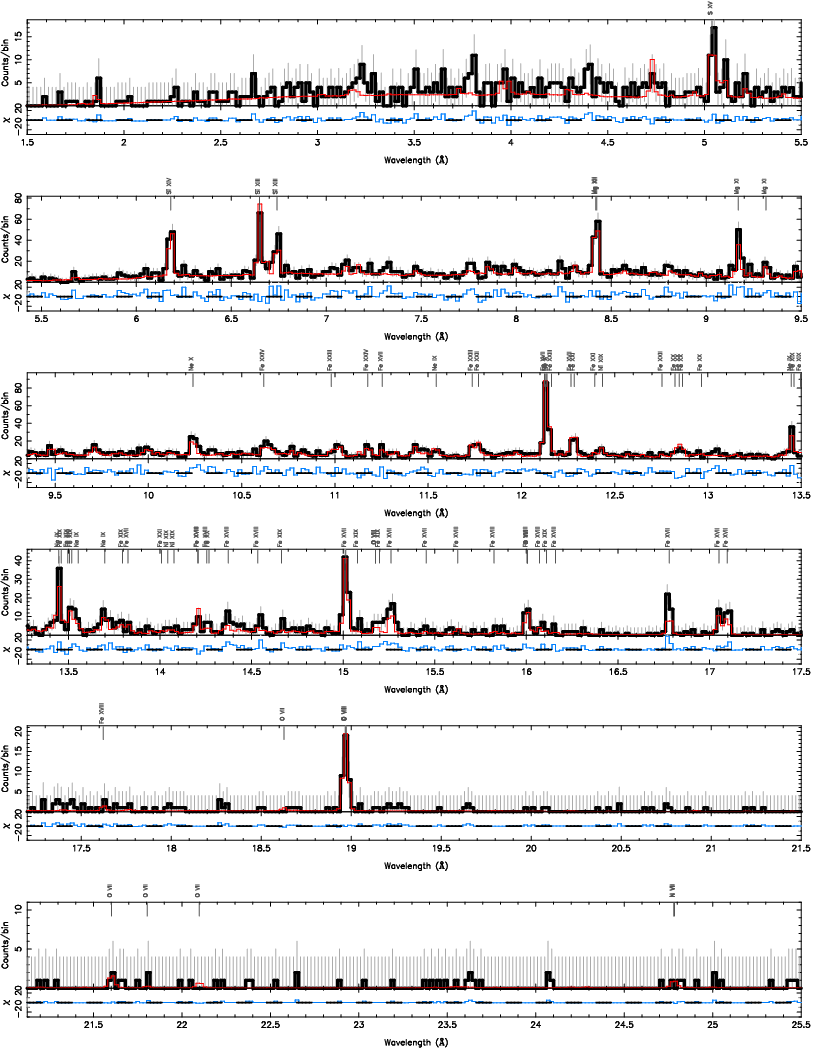,width=17cm}}
\caption{Comparison of the \heg+\meg\ spectrum of \hd-A with the 
	model (light histogram) assuming \nh$= 1.5 \times 10^{21}$~cm$^{-2}$. 
	Residuals are also shown.
       \label{fig:specA}}
\end{figure*}

We reconstructed an emission measure distribution ($EMD$) and
elemental abundances by minimizing line flux residuals for features
marked in Table~\ref{tab:linefluxes} with an ``E'' in the ``Use''
column.  This was an iterative process, since some features are
blended or mis-identified: we excluded those with very large residuals
during the fitting process.  We also excluded the He-like resonance
and intercombination lines since they are density dependent.
Theoretical emissivities from APED were integrated over the
temperature range of sensitivity after weighting by the trial $EMD$
and abundances.  The fit was subject to the constraint that the $EMD$
be smooth.  To obtain some estimate of the uncertainty in the
reconstructed $EMD$ and abundances, we performed 100 Monte-Carlo
iterations in which the observed fluxes were randomly perturbed using
a Gaussian distribution with their measurement uncertainties as the
Gaussian $\sigma$.  We computed the standard deviation in the
resulting $EMD$ at each temperature and for the abundances and use
these for the fit uncertainty.  This is not rigorously correct because
it does not account for correlations in the allowed solutions.  Other
examples of this method, and further details, can be found in
\citet{Huenemoerder06, Huenemoerder07}, along with requisite caveats and
limitations of emission measure modeling.   As a final check, the
synthetic spectrum was evaluated and compared to the observation (see
Figure~\ref{fig:specA}).  Model line fluxes are listed in
Table~\ref{tab:linefluxes}, the $EMD$ is shown in
Figures~\ref{fig:dem}, and abundances are given in
Table~\ref{tab:abund}.

The emission measure distribution shown in Figure~\ref{fig:dem} is
characterized by significant amount of plasma over a wide temperature
range, as anticipated on the basis of a visual inspection of the
spectrum, and by a peak at $\log T[K] \sim 6.9$.  These
characteristics are similar to the temperature distributions derived
for several coronae of late-type stars (see e.g., \citealt{Sanz03a}),
typically presenting an emission measure peak around $10^7$~K.
However, coronae of active stars with X-ray luminosity $\gtrsim
10^{30}$erg~s$^{-1}$ often present an additional significant peak at
higher temperatures (a few $10^{7}$~K) typically associated with
flaring activity.  Hot plasma is present in \hd-A as well, but with a
lower relative weight than observed in other stellar coronae at a
similar activity level.

\begin{figure*}[!tbhp]
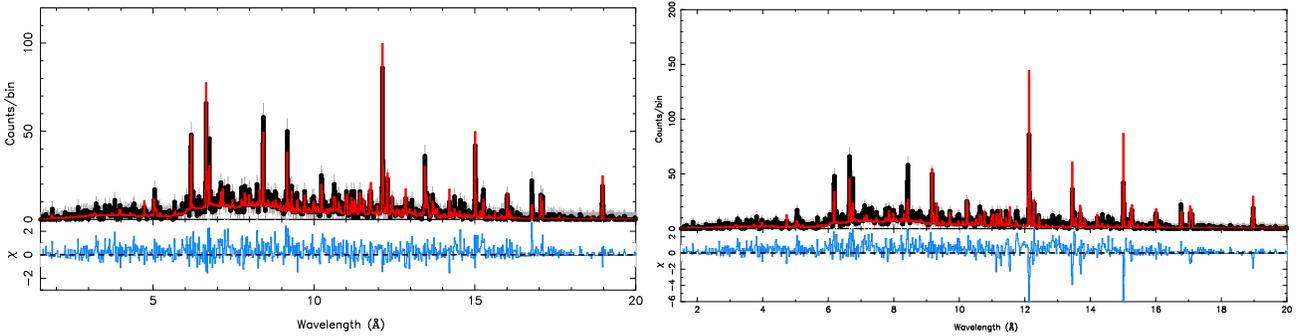

\centerline{\psfig{figure=f11a.ps,width=8.5cm}
	    \psfig{figure=f11b.ps,width=8.5cm}}
\caption{Comparison of the \heg+\meg\ spectrum of \hd-A with
	the spectrum simulated using the $EMD$ model ({\em left})
	and the $2T$ model ({\em right}). 
	Note the different scales on $\chi$.
    \label{fig:specC}}
\end{figure*}

In Figure~\ref{fig:dem} we also compare the $EMD$ model with the $2T$
model obtained from the low-resolution spectra
(model I in Table~\ref{tab:fit_zoA}): the emission measure
values for the components of the $2T$ model are scaled for comparison
purposes, but their sum compares well with the integrated emission
measure distribution.  Figure~\ref{fig:specA} shows the synthetic
spectrum from the $EMD$ model, compared with the actual spectrum.
Figure~\ref{fig:specC} compares the predicted spectrum from the two
models with the actual observed spectrum and shows that while the
overall characteristics of the $2T$ model are somewhat similar to the
detailed $EMD$ model, the latter reproduces the spectral features of
the high-resolution spectrum much more satisfactorily.

\begin{deluxetable}{lccc}
 \tabletypesize{\footnotesize}
 \tablecaption{Element abundances of X-ray emitting plasma from high 
 		resolution spectrum. \label{tab:abund} }
 \tablewidth{0pt}
 \tablehead{
   \colhead{Element} &  \colhead{$A/A_\odot$\tablenotemark{a}} & 
   \colhead{$A/A_{\rm phot}$\tablenotemark{b}}  & 
   \colhead{$A_{\rm phot}/A_\odot$~[dex] \tablenotemark{c}}
 }
 \startdata
O    &  $0.65 \pm 0.14$  &  $0.82 \pm 0.33$                 &  $-0.10 \pm 0.14$  \\
Ne   &  $1.05 \pm 0.10$  &  ...                             &  ...  \\
Mg   &  $0.66 \pm 0.08$  &  $1.02 \pm 0.20$                 &  $-0.19 \pm 0.07$  \\
Si   &  $0.91 \pm 0.10$  &  $0.87 \pm 0.25 / 0.45 \pm 0.11$ &  $0.02 \pm 0.11 / 0.31 \pm 0.09$  \\
S    &  $1.24 \pm 0.36$  &  $1.75 \pm 0.80$                 &  $-0.15 \pm 0.14$  \\
Fe   &  $0.50 \pm 0.05$  &  $0.42 \pm 0.19$                 &  $0.08 \pm 0.18$  \\
\enddata
\tablenotetext{a}{Element abundances of the X-ray emitting plasma
		from the high resolution X-ray spectrum, expressed as 
		relative to solar \citep{GrevesseSauval}. We also
		list 1$\sigma$ errors.}
\tablenotetext{b}{Element abundances of the X-ray emitting plasma
		from the high resolution X-ray spectrum, expressed as
		relative to photospheric abundances. Errors are 
		calculated propagating the errors on both the 
		photospheric abundances, and the abundances of the 
		X-ray emitting plasma.}
\tablenotetext{c}{Photospheric values of element abundances from 
		\cite{Acke04}; for Si we list their values derived
		from Si\,{\sc i}, and from Si\,{\sc ii}	respectively.}
\end{deluxetable}

The element abundances of the X-ray emitting plasma of \hd-A
derived from the analysis of its \hetgs\ spectrum are listed in 
Table~\ref{tab:abund}, both compared to solar values 
\citep{GrevesseSauval} and to the photospheric values of \hd-A
from \cite{Acke04}. 
These abundances show a general depletion of elements in the 
X-ray emitting plasma with the exception of S, generally in line
with X-ray studies of other PMS stars (e.g., \citealt{Maggio07}), 
however with some possible differences: the Ne/Fe abundance 
ratio is relatively low ($\sim 2$),with respect to the extremely 
high values observed in other young stars (see e.g., 
\citealt{Kastner02,Argiroffi05} finding values of Ne/Fe $\sim 10$ for 
PMS stars in TWA); also, possible dependence of fractionation 
on the element first ionization potential, as often observed
in coronae of late-type stars (e.g., \citealt{Telleschi05}),
is not obviously present in the X-ray emitting plasma of \hd-A.

\subsubsection{Density and Temperature Diagnostics from Line Ratios}

The He-like triplets, in particular the cooler \neix\ and \ovii\ 
triplets, thanks to their sensitivity to plasma density and 
temperature (e.g., \citealt{Gabriel69}), and to photoexcitation from 
possible UV field, provide effective diagnostics for the X-ray 
emission mechanism in young stars (e.g., \citealt{Kastner02}). 
In particular, in absence of intense UV field, the ratio of the 
forbidden to intercombination line, $R \equiv f/i$, is sensitive to plasma 
densities above a critical density value that depends on the element 
and increases from approximately $3 \times 10^{10}$~cm$^{-3}$ for 
\ovii\ to about $3 \times 10^{13}$~cm$^{-3}$ for \sixiii. 
The presence of strong UV field, either photospheric for hot stars 
or produced in accretion shocks in young stars, affects the $R$ ratio 
similarly to high densities, i.e.\ decreasing its value. 
In the case of \hd-A the stellar effective temperature is low enough
(see Table~\ref{tab:Hae}) to rule out any effect of the photospheric 
UV field on the X-ray triplet lines.
The ratio of the summed forbidden and intercombination lines with the 
resonance line, $G \equiv (f+i)/r$, provides instead a temperature diagnostic
(though with some caveats on the reliability of the diagnostics, as discussed
e.g., by \citealt{Testa04b}).

The \ovii\ He-like triplet at 22\AA\ is too weak in \hetgs\ spectrum 
of \hd-A to provide any meaningful diagnostic (only 3 counts are 
detected for the strongest triplet line, the 21.6\AA\ resonance line).
\neix, \mgxi, and \sixiii\ have good signal-to-noise ratios and 
provide useful info on the plasma density and temperature.
The \sixiii\ triplet is at the low density limit as for all other observed
high resolution spectra of late-type stars (e.g., \citealt{Testa04b}).

\begin{figure*}[!tbhp]
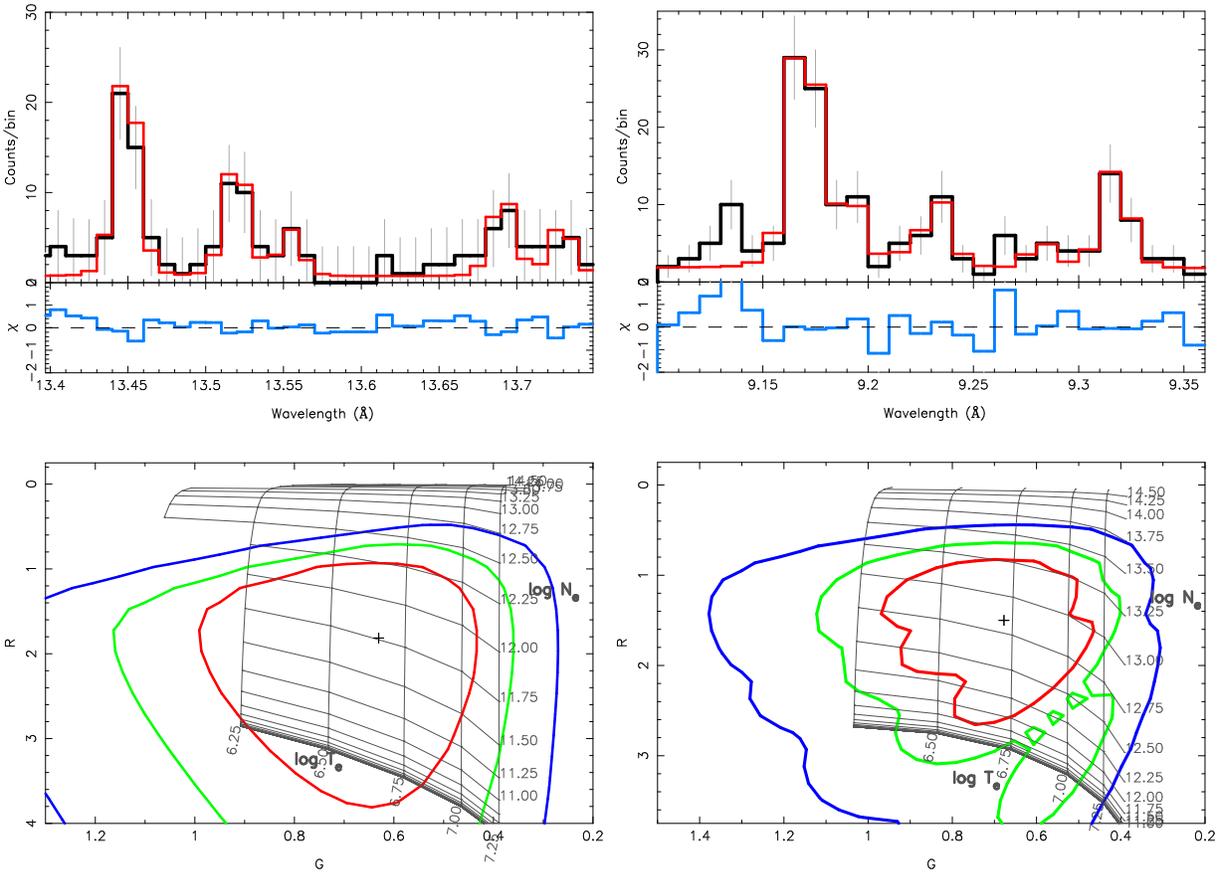

\centerline{\psfig{figure=f12a.ps,width=8cm}
	    \psfig{figure=f12b.ps,width=8cm}}
\caption{\cha-\hetgs\ spectrum of \hd-A in the \neix\ ({\em left})
	and in the \mgxi\ ({\em right}) He-like triplet wavelength 
	regions (black; rebinned by a factor 2 with respect to 
	original wavelength grid) with best fit model (red).
	The lower panels show the corresponding confidence contours 
	(68\%. 90\%, 98\%) in the $R-G$ parameter space and indicate
	the density and temperature diagnostics derived from the 
	triplet line ratios.
       \label{fig:cNe9}}
\end{figure*}

In Figure~\ref{fig:cNe9} we show the \neix\ and \mgxi\ triplet regions 
with our best fit parametric models (upper panels), and the confidence 
contours in the parameter space of the line ratios $R$ and $G$ (lower
panels). 
In the $R$ and $G$ parameter space we overplot the grid of corresponding 
densities and temperature according to \cite{Smith01}.
The fitting model for the \neix\ triplet region includes the blends 
of mostly Fe lines as computed from the $EMD$ model (see e.g., 
\citealt{Huenemoerder06}).
The best fit to the \neix\ triplet lines corresponds to a density
$n_{\rm e} \sim 6 \times 10^{11}$~cm$^{-3}$; however the relatively
low statistics of the spectrum does not allow to put stringent limits
on this value and we can reliably derive only an upper limit to the 
density of about $n_{\rm e} \lesssim 6 \times 10^{12}$~cm$^{-3}$.

The fits to the \mgxi\ triplet region included \nex\ Ly series blends,
and blends of Fe lines (see e.g., \citealt{Testa04b}).
\mgxi\ allows to put slightly better constraints on the plasma 
density: the best fit density is $n_{\rm e} \sim 7 \times 10^{12}$~cm$^{-3}$
with a $1 \sigma$ interval $1-18 \times 10^{12}$~cm$^{-3}$,
somewhat high compared with typical values of active late-type 
stars (e.g., \citealt{Testa04b}).  Also the $G$ ratio is somewhat 
extreme when compared with typical coronal values, being at the 
low-end of the range of measured ratios \citep{Testa04b},
even though the rather large uncertainties do not allow to draw
definite conclusions. Interpreting the $G$ ratio as a temperature
diagnostics would point to a plasma temperature in agreement with the
expected values for the formation of these lines. This is at variance
with findings relative to He-like triplets of stellar coronae whose 
$G$ ratios generally underestimate the plasma temperature.
Therefore it is difficult to interpret an unusually low $G$ ratio
in \hd-A as a temperature effect considering that its temperature
structure does not seem peculiar when compared to stellar coronae.
We note that also the \neix\ $G$ ratio has a value among the lowest
measured in a large sample of active stars at different activity levels
\citep{Ness04}.

\subsubsection{Line profiles}
\label{s:line_prof}

Spectral line profiles in X-rays allow us to probe the bulk velocity
of hot plasma and its turbulent state.  We carried out the measurement
of line centroid positions of several prominent emission lines
detected in the \meg\ spectra with sufficient counts for meaningful
fits. 

The methodology used here is somewhat different from the line-flux
measurements described in \S\ref{s:hetgspec}.  Line centroids are
very sensitive to blending, even with weak features.  Here we take
advantage of a model spectrum which predicts the relative positions
and relative strengths of many features in any region.  Including
thermal broadening in the flux evaluation automatically accounts for
any line-width difference between species, and turbulent broadening
can be included to test for excess width.  For the best accuracy,
these fits were only done on strong-line regions, whereas flux
measures for emission measure reconstruction were also done for weak
lines, since the absence of a line can be significant for the
temperature structure.  Since the line widths are barely resolved,
if at all, the width measurement can best be done on only the
highest signal-to-noise features, and though the \heg\ has higher
spectral resolution, the higher counts in the \meg\ can makes its
determination of width more significant.

Utilizing the latest calibration product for \meg\ line spread
function (available for \cha\ CALDB v.3.4.0 or higher), grating ARFs
and RMFs are generated for \meg\ +1 and -1 orders. Then the $ \pm 1$
order \meg\ spectra are simultaneously fit with a simple 2T model
similar to the ones listed in Table~\ref{tab:fit_zoA} (close to model
I) with line centroid and broadening parameters as free parameters. We
have taken thermal broadening into account upon fitting. Since the
observations were taken at four different times over the two-month
period, each spectrum was corrected for appropriate barycentric motion
of the Earth (-8.3, -9.9, -9.4, and +13~km~s$^{-1}$ for ObsID 6444,
7319, 7320 and 7326, respectively) prior to the fitting process.

\begin{deluxetable}{lcc}
 \tabletypesize{\footnotesize}
 \tablecaption{Velocity from Line Shifts Measurements \label{tab:lineprofiles} }
 \tablewidth{0pt}
 \tablehead{
   \colhead{Ion}  &  \colhead{$\lambda$ [\AA]} &  
   \colhead{$v_{\rm rad}$ [km~s$^{-1}$]\tablenotemark{a}}  }
 \startdata
\sixiv\    & 6.183  &  0.0 [$-140,+240$] \\
\mgxii\    & 8.422  &  76  [$ -61,+220$] \\
\mgxi\     & 9.169  &  120 [$-100,+220$] \\
\nex\      & 12.135 &  0.0 [$ -96,+85$] \\
\fexvii\   & 15.014 &  43  [$-52,+100$] \\
\fexvii\   & 17.096 &  41  [$-37,+110$] \\
\oviii\    & 18.970 &  43  [$-31,+130$] \\
\enddata
\tablenotetext{a}{The 99\% confidence intervals are listed in square brackets.}
\end{deluxetable}

We selected eight prominent lines in a wavelength range from
6 to 20\AA: the H-like Ly$\alpha$ transitions of Si, Mg, Ne, and
O, and the \fexvii\ transitions at $\sim$ 15.0 and 17.1\AA.  
The measured line velocities are tabulated in Table \ref{tab:lineprofiles}.

The mean line centroid velocity of the source is $46 \pm 42$ km~s$^{-1}$
with a 99\% confidence level. The large uncertainty in the measurement
results from the low S/N level detected in each emission lines.
As noted, \hd-A is a binary system. Generally a small radial
velocity correction should be made -- prior to line fitting --
for a spectrum obtained at each epoch. However, we intentionally
neglect this since (i) we do not know the relative contribution of
each component to the spectra and (ii) all of the spectra were
obtained in the phase ranging from 0.25 -- 0.5 (see bottom panel of 
Figure~\ref{fig:lcHETG_phf}), placing the stars
near apastron where the change in radial velocity happens to be
very small, of order of 7~km~s$^{-1}$ based on \citealt{Bohm04}.

At the observed times, the radial velocity of the primary star, the 
HAe star, should be observed at $\approx +20$~km~s$^{-1}$, whereas 
the companion, K3, should be at $\approx +10$~km~s$^{-1}$. 
While the measured radial velocity prefers the primary star to be 
the source of hard X-rays, the companion is not statistically ruled 
out as the source.
Furthermore, with the radial velocity correction of the systemic binary
motion, the measured line centroid velocity becomes consistent with 
the hard X-ray source being at rest.

We also measured the width of the lines, using \meg\ $\pm 1$
(and \heg\ $\pm 1$ for lines with sufficient counts like the \nex\ 
Ly$\alpha$ transition) order spectra. 
The line widths are consistent with zero broadening for all the 
strongest lines, with upper limit for $v_{\rm turb}$ of the
order of 300~km~s$^{-1}$.  The fit to the \fexvii\ 15\AA\ line 
provides a best fit width corresponding to a turbulent velocity
$v_{\rm turb}$ of about 280~km~s$^{-1}$, and barely compatible with 
0 at the 99\% level. However, we note that the statistics are limited, 
and a small error in the continuum can cause $v_{\rm turb}$ to be 
compatible with 0 at the 68\% confidence level. Also, the other 
\fexvii\ lines around 17\AA\ are compatible with zero broadening 
casting doubts on the results to the fit to the 15\AA\ line.

\section{Discussion}
\label{s:discuss}

The \hd\ young association, with presumably coeval stars spanning 
0.15-2.3\msol, offers a laboratory for studying X-ray 
emission in young low/intermediate mass stars. 
The brightest X-ray source (and also so in optical, UV, and IR bands)
in the stellar group is a system harboring a binary of intermediate 
mass stars (2.25/1.75 \msol) with a HAe star, and a slightly lower 
mass star in an early evolutionary stage with spectral type K3. 
The Herbig Ae star \hd-A with its proximity, high \lx, and low 
line-of-sight extinction is one of the few HAe stars accessible
to high resolution X-ray spectroscopy, allowing us to investigate
the nature of the X-ray emission from young intermediate mass stars 
that is still a mystery. The age of the binary system is estimated 
between 2 and 5~Myr (\citealt{Bohm04}, \FHa), and the HAe star is 
correspondingly in the late stages of its pre-main sequence phase.

The long \cha-\hetgs\ observation provides important clues to the 
X-ray emission mechanism through spectral diagnostics and 
variability.
Different scenarios may be considered to explain the X-ray emission 
of \hd-A: (1) all X-ray emission comes from the HAe star; (2) all 
X-ray emission comes from the companion; (3) both binary components 
contribute to the observed X-ray emission.
The companion hypothesis is still considered a viable explanation 
for the X-ray emission of Herbig Ae stars, considering that these 
intermediate mass stars are not expected to produce X-rays either 
through wind shocks, as in earlier type stars, because their winds 
are weak, or through magnetic activity as in later type stars, 
because they should not have an efficient solar-like dynamo mechanism 
producing magnetic fields since they lack sizable convective zones.
However, some alternative mechanisms might be at work in the early 
stages of their evolution: (i) X-ray emission from accretion shocks,
as in some classical T Tauri stars (e.g., \citealt{Kastner02}), 
(ii) coronal activity due to non-solar dynamo mechanism 
\citep{Tout95}, (iii) magnetically confined winds \citep{Babel97}.
All these mechanisms require the presence of some sizable magnetic 
field in this evolutionary stage. 
The magnetic field in HAe stars has been investigated on large sample 
of these intermediate mass stars, and a small number of them do show 
evidence of magnetic fields of strength from a few tens of Gauss to 
a few hundreds of Gauss in a few cases 
\citep{Donati97,Hubrig04,Wade05,Wade07,Hubrig07,Catala07}.
\hd-A is one of those showing that, whatever the mechanism that generates 
it, at some stage these intermediate mass stars can have magnetic fields.
The expected X-ray properties are different depending on the mechanisms
with (i) characterized by soft X-rays emitted by high density plasma, 
and harder X-ray emission due to magnetic confinement (and presumably 
heating) in (ii), and (iii). In particular in the scenario of coronal 
activity we expect a large degree of variability and flare-like dynamic 
events.

What do we know from observations of other HAe stars?
X-ray observations with \cha\ allow to resolve the Herbig Ae stars 
from close ($\gtrsim 1$\arcsec) companions and therefore reduce the 
confusion, and still indicates that a significant fraction 
($\sim 35$\%; \citealt{Stelzer06}) of Herbig Ae star with unknown 
companions are X-ray sources.
As shown by \cite{Stelzer06}, the HAe stars have X-ray properties
(\lx, $T$) very similar to lower mass young stars; therefore they
conclude that either their emission and, likely, X-ray 
production mechanisms are very similar to TTS (i.e., coronal), or 
their X-ray emission comes from yet to be discovered companions.
A few exceptions exist: HD~163296 has been observed with \cha\ 
and shows an unusually soft spectrum, that \cite{Swartz05} tentatively
attribute to accretion shocks as in some CTTS; analogously AB~Aur 
shows a very soft spectrum \citep{Telleschi07}.

\begin{figure*}[!tbhp]
\centerline{\psfig{figure=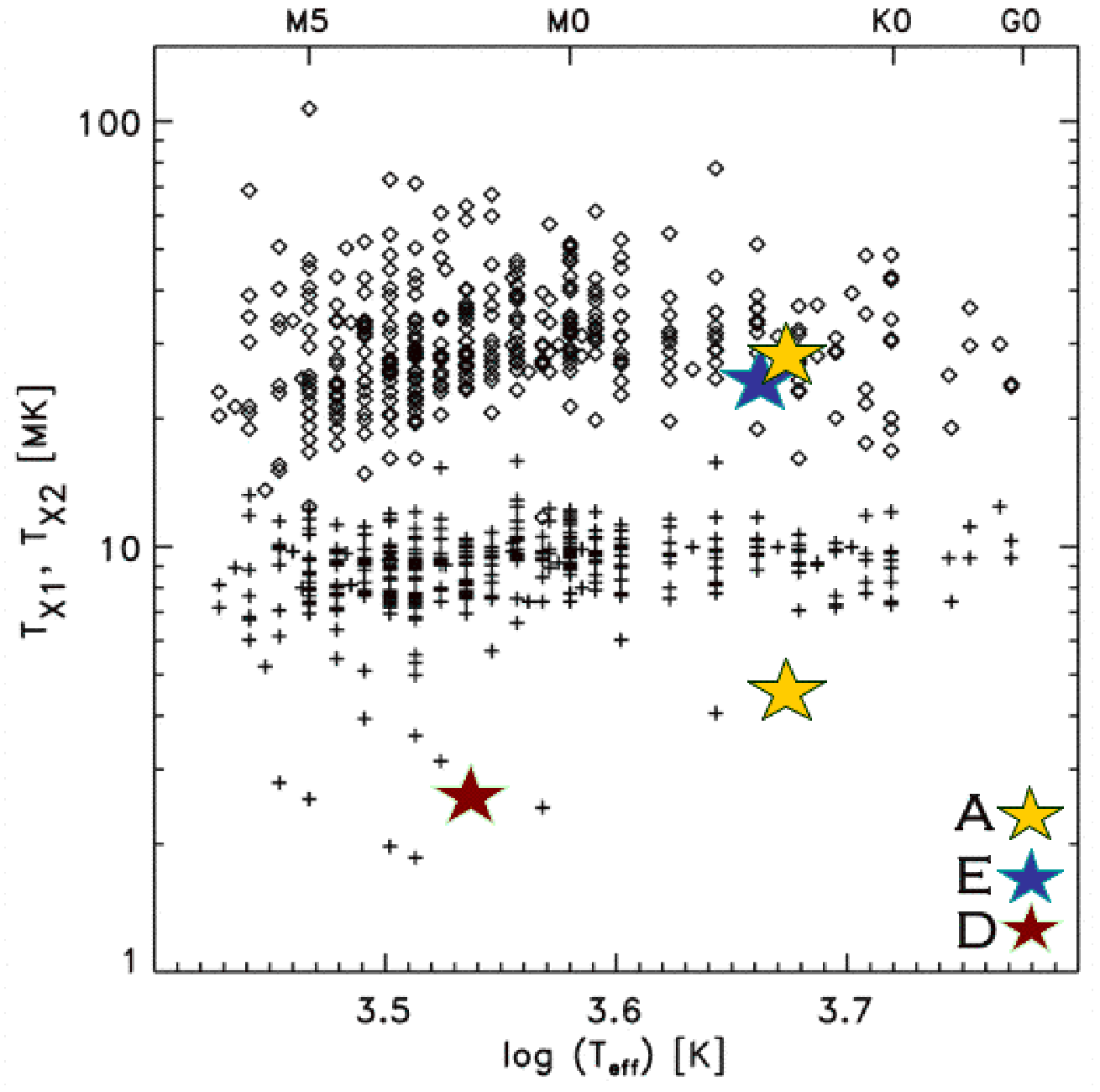,width=7.6cm}
	   \psfig{figure=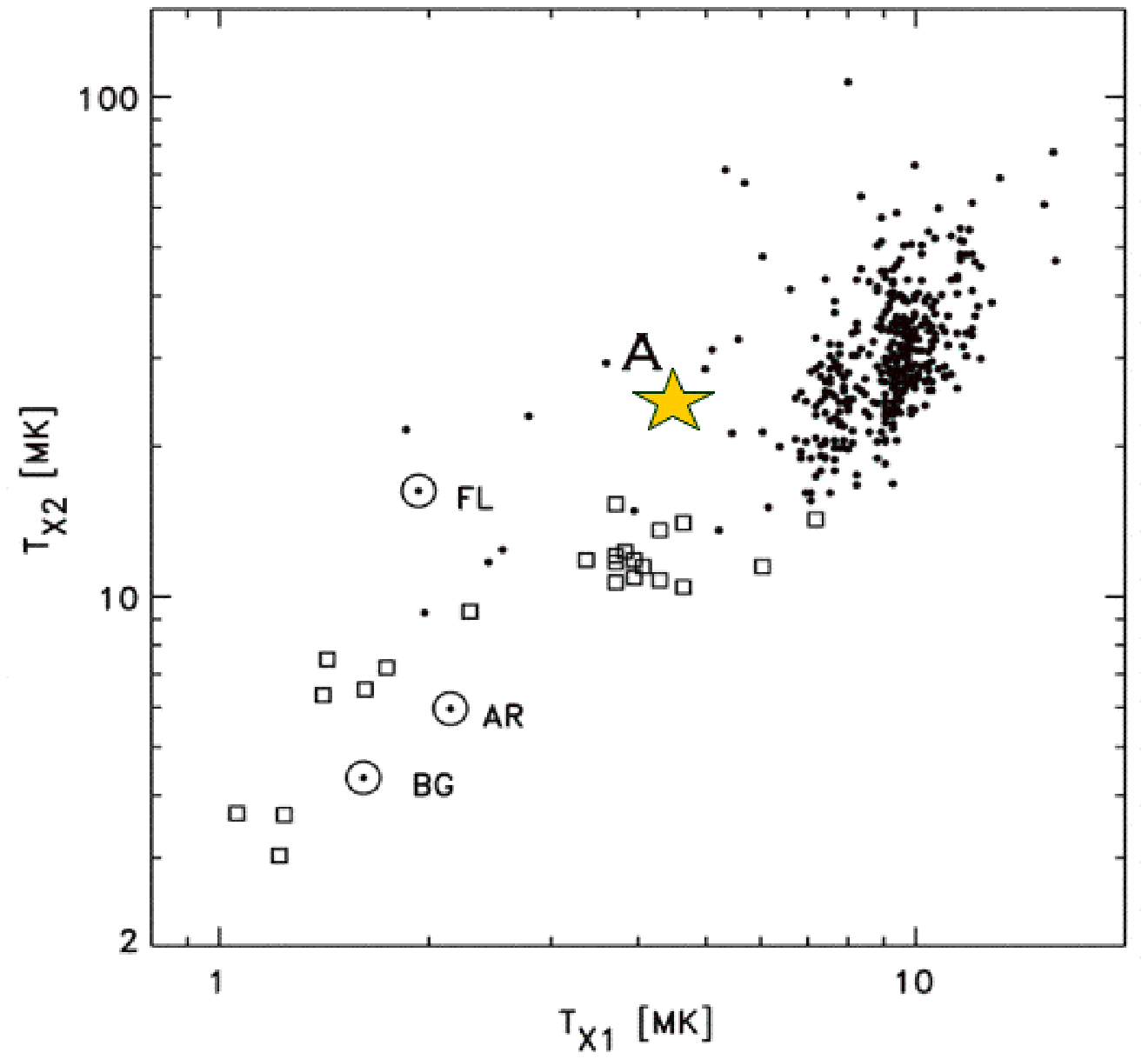,width=8cm}}
\caption{Comparison of plasma temperature of the \hd\ members, obtained
	from 2-T (for source A) or 1-T (for source D, and E) analysis 
	of their zero-order spectra, with results from the 
	COUP sample \citep{Preibisch05}. We used plots from 
	\cite{Preibisch05} (their Fig.~12, and 13; reproduced by 
	permission of the AAS) and superimposed 
	the values found for \hd-A, D and E (star symbols of different
	colors: A = yellow, D = red, and E = blue). {\em Left panel:} 
	temperature of X-ray emitting plasma vs.\ effective temperature.
	$T_{\rm X1}$ ({\em crosses}), and $T_{\rm X2}$ ({\em diamonds}) 
	are the two temperature values of a 2-T model fit to the 
	low resolution X-ray spectra.
	{\em Right panel:} $T_{\rm X1}$ vs.\ $T_{\rm X2}$.
	\cite{Preibisch05} (see references therein) include in the 
	plot results for a sample of G- and K-type MS stars ({\em open
	squares}), and typical values for structures in the solar 
	corona (BG = background corona; AR = active region; FL = flares).
	\label{fig:comp_COUP}}
\end{figure*}

The spectroscopic companion to \hd-A is a later spectral type star 
(similar to SU~Aur) and similar stars are known to be strong X-ray 
emitters (see e.g., \citealt{Getman05}, and \citealt{Gudel07}). 
Therefore the K3 companion is expected to contribute significantly 
to the overall X-ray emission. 
We explore the companion hypothesis assuming that the X-ray emission 
is entirely due to the K-type companion to \hd-A, and comparing its
properties with the X-ray properties of stars of similar mass and 
evolutionary stage.
The \cha\ Orion Ultradeep Project (COUP; \citealt{Getman05}) provides
information for putting the X-ray emission properties of the \hd\ 
members into context.
In Figure~\ref{fig:comp_COUP} we compare some spectral properties of 
the \hd\ members with the large sample from COUP \citep{Preibisch05}.
In particular \cite{Preibisch05} analyzed the \cha-\acis\ spectra 
of the COUP sources with a two-temperature model and studied how
the best fit temperatures change with spectral type. We superimpose
on the results from COUP our temperature values from the modeling 
of the zero-order spectra, assuming a spectral type K3 for \hd-A
(Figure~\ref{fig:comp_COUP}, left panel).
We analyze in the same context also the results found for source D and
E, which have well determined spectral types. 

The plots of Figure~\ref{fig:comp_COUP} show that the hotter of the 
two temperatures fitting the spectrum of source A compares well 
with the high temperature component found for early K-type stars
in the young star forming region of Orion, and also with the 
temperature of source E which has same spectral type as the 
spectroscopic binary companion to A. 
On the other hand, the soft component seems significantly cooler 
than the soft component of these similar stars.
This soft component is reminiscent of the soft spectra found for the
two Herbig Ae stars, HD~163296 and AB~Aur \citep{Swartz05,Telleschi07}.
These recent findings might lend support to the idea that 
characteristic soft X-ray emission could be a defining property 
of X-ray spectra of HAe stars.
The persistent hot component, however, cannot be explained by 
either the wind models or magnetic infall scenario proposed for the 
CTTS, and therefore the detection of the hot component suggests the
presence of magnetically confined plasma. This hot emission is likely 
produced, at least partially, by the later type companion; on the 
other hand the X-ray emission mechanisms -- emission from magnetically 
confined coronal plasmas, or from accretion streams, or from magnetically
confined winds -- may not be mutually exclusive and may all contribute
to the X-ray emission in some HAe systems.
Possible generation of magnetic fields in \hd-A cannot be easily 
explained with existing models. The estimated position in the HR 
diagram and evolutionary tracks imply that \hd-A is already 
completely radiative and therefore it should not be able to produce 
any magnetic field by solar-like dynamo mechanism. In the alternative
shear magnetic model by \cite{Tout95}, X-ray luminosities at levels
high enough to explain the observed \lx\ of HAe stars can be produced
for time scales of the order of 1~Myr, and then they are rapidly 
decreasing as $L_{\rm X} (t) = L_{\rm X_0} (1+ t/t_0)^{-3}$, where
$L_{\rm X_0}$ depends on the stellar parameters. As calculated by
\cite{Skinner04} depending on the set of stellar parameters adopted
in the ranges discussed in \S\ref{s:targets} this model underestimates
the observed \lx\ by a factor $\sim 4$ (for spectral type A4\,{\sc iv}e
and age $t \sim 2$~Myr) to $\sim 25$ (for the more recently determined
spectral type A7.5\,{\sc v}e-A8\,{\sc v}e and age 2-5~Myr).

HD~163296, AB~Aur and \hd-A show similar level of X-ray emission with
fractional luminosity $\log$\lx/\lbol\ of about -5.5 \citep{Swartz05},
-5.6 \citep{Telleschi07}, and -5 (this work; $\log$\lx/\lbol$\sim -4$
if we assume that the K3 companion is the X-ray source)
respectively.
Another similarity between \hd-A and AB~Aur consists in their X-ray 
variability: both sources are slowly varying within a factor $\sim 2$, 
do not present any clear evidence of flaring activity, and their X-ray 
lightcurve is compatible with their respective rotational period as 
derived from periodicity of optical lines. 
The \hetgs\ observations presented in this paper do not provide enough
phase coverage to definitely determine the presence of X-ray modulation
on the timescales of modulation of the H$\alpha$ line, however, if the 
observed modulation is confirmed to be consistent with the rotation 
period of the HAe star, this would imply that a significant portion
of the observed X-ray emission is produced by the HAe star.
X-ray rotational modulation has been previously observed in a number of
X-ray bright PMS stars, such as for instance by \citet{Flaccomio05} 
who found evidence of X-ray emission being modulated by rotation in
at least 10\% of the Orion PMS stars sample observed in the COUP campaign;
these findings are also compatible with theoretical modeling of TTS 
coronae from extrapolations of surface magnetograms derived from 
Zeeman-Doppler imaging \citep{Gregory06}. 

The high resolution \hetgs\ spectra offer additional spectral 
diagnostics providing clues to the yet unestablished X-ray 
production mechanisms in young intermediate mass stars. 
Particularly useful diagnostics are provided by the He-like triplets:
specifically the \neix\ and \mgxi\ triplets, since the \ovii\ triplet 
is essentially undetected, and the \sixiii\ lines are at the low 
density limit (therefore providing only an upper limit to density of 
a few $10^{13}$~cm$^{-3}$).
Both Ne and Mg He-like triplets line ratios indicate plasma densities
somewhat larger than typically observed in stellar coronae at the same
level of activity, though the limited statistics do not provide 
very stringent constraints to these findings.
In the scenario tentatively drawn above, with the Herbig Ae star
being the main contributor to the cool ($< 10^{7}$~K) emission and the
K3 companion producing most of the hard emission, these triplets 
lines would be produced mainly by the Herbig Ae star. Considering
the presence of significant ongoing accretion (at an estimated
rate of $\sim 10^{-8}$\msol~yr$^{-1}$), in analogy with TW~Hya
and other accreting T~Tauri stars showing high density in their
high resolution spectra (e.g., 
\citealt{Kastner02,Stelzer04,Schmitt05,Gunther06,Argiroffi07}),
the high densities observed for \hd-A can be interpreted as 
a signature of X-ray production in accretion shocks.
The high resolution spectra also provide line profile diagnostics
potentially able to determine the contribution to the X-ray emission
of each component of a binary system (see e.g., \citealt{Ishibashi06}).
We find that the measured line centroids for the strongest lines
(see \S\ref{s:line_prof}) indicate that the Herbig Ae star is more
likely, with respect to the K3 companion, to be the primary X-ray 
source.
However, in our case the involved velocities are small, and the 
statistics limited, therefore not providing conclusive results.
As discussed in \S\ref{s:line_prof} no convincing evidence
of line broadening is found in the strong X-ray lines.

Among the other sources in the association, source D shows interesting
characteristics: as clear from Figure~\ref{fig:comp_COUP} (left panel), 
source D is characterized by an unusually cool plasma temperature with 
respect to the bulk of similar M-type young stars in Orion. 
Even though no clear evidence of ongoing accretion is available for
this star, its ${\rm H} \alpha$ emission is near the CTTS limit
indicating that this star might still be actively accreting from its
circumstellar disk; in this scenario, the soft excess might indicate
that X-rays produced by shocked accreting plasma dominate the X-ray 
emission of this star, analogously to the peculiar case of TW~Hydrae
\citep{Kastner02}, and differently from the majority of TTS where the 
coronal component generally dominates the overall X-ray emission
(e.g., \citealt{Preibisch05}).

\section{Conclusions}
\label{s:conclusions}
We have analyzed \cha-\hetg\ observations of young low-mass and 
intermediate-mass stars in the small stellar group associated with
the Herbig Ae star \hd\ to investigate the X-ray emission 
mechanisms in their pre-main sequence phase.  
The close ($\sim 15$~AU) binary system of the HAe star and its K-type 
companion cannot be resolved in X-rays, however our findings with 
respect to the X-ray variability and spectral properties provide 
clues to the origin of the X-ray emission and the X-ray production 
mechanisms.
The modulation on time scales of the estimated rotation period of the
HAe star suggests that the primary contributes significantly to the
observed X-ray emission.  The spectral analysis reveals a strong 
soft component ($\sim 0.5$~keV) significantly cooler than typically
found in K-type stars of evolutionary stage similar to that of the 
\hd-A companion ($0.7-1$~keV). We interpret these findings in a possible
scenario where the soft X-rays are produced mainly by the HAe star and 
the hard emission by the later-type companion. 
The analysis of the \hetgs\ spectrum provides additional diagnostics,
in particular through the He-like triplets: the line ratios of the 
\mgxi\ and \neix\ triplets suggest the presence of plasma at high 
densities of about $10^{12}$~cm$^{-3}$, possibly indicating accretion 
related X-ray production mechanism.

The observed X-ray emission of the other sources is typical of pre-main
sequence stars of similar ages and spectral types, except for 
the M-type T Tauri star \hd-D for which we find evidence of 
remarkably soft emission, with temperature of 3~MK reminiscent of
the X-ray emission of the classical T Tauri star TW~Hya whose
emission is attributed to shocks in the accreting plasma.

\begin{acknowledgements}
This research was supported by SAO contract SV3-73016 to MIT for support 
of the {\em Chandra X-ray Center}, which is operated by SAO for and on 
behalf of NASA under contract NAS8-03060.  This research was based on 
Chandra Cycle 7 Guaranteed Time Observation program selected by Claude 
Canizares, whom we thank for providing the opportunity to analyze 
these data, and for useful discussions and comments on this manuscript.
\end{acknowledgements}

\bibliographystyle{apj}

\begin{thebibliography}{68}
\expandafter\ifx\csname natexlab\endcsname\relax\def\natexlab#1{#1}\fi

\bibitem[{{Acke} {et~al.}(2005){Acke}, {van den Ancker}, \&
  {Dullemond}}]{Acke05}
{Acke}, B., {van den Ancker}, M.~E., \& {Dullemond}, C.~P. 2005, \aap, 436, 209

\bibitem[{{Acke} \& {Waelkens}(2004)}]{Acke04}
{Acke}, B., \& {Waelkens}, C. 2004, \aap, 427, 1009

\bibitem[{{Alcala} {et~al.}(1995){Alcala}, {Krautter}, {Schmitt}, {Covino},
  {Wichmann}, \& {Mundt}}]{Alcala95}
{Alcala}, J.~M., {Krautter}, J., {Schmitt}, J.~H.~M.~M., {Covino}, E.,
  {Wichmann}, R., \& {Mundt}, R. 1995, \aaps, 114, 109

\bibitem[{{Anders} \& {Grevesse}(1989)}]{Anders89}
{Anders}, E., \& {Grevesse}, N. 1989, \gca, 53, 197

\bibitem[{{Appenzeller}(1994)}]{Appenzeller94}
{Appenzeller}, I. 1994, in Astronomical Society of the Pacific Conference
  Series, Vol.~62, The Nature and Evolutionary Status of Herbig Ae/Be Stars,
  ed. P.~S. {The}, M.~R. {Perez}, \& E.~P.~J. {van den Heuvel}, 12--+

\bibitem[{{Argiroffi} {et~al.}(2007){Argiroffi}, {Maggio}, \&
  {Peres}}]{Argiroffi07}
{Argiroffi}, C., {Maggio}, A., \& {Peres}, G. 2007, \aap, 465, L5

\bibitem[{{Argiroffi} {et~al.}(2005){Argiroffi}, {Maggio}, {Peres},
  {Stelzer}, \& {Neuh{\"a}user}}]{Argiroffi05}
{Argiroffi}, C., {Maggio}, A., {Peres}, G., {Stelzer}, B. \&
  {Neuh{\"a}user}, R. 2007, \aap, 439, 1149

\bibitem[{{Babel} \& {Montmerle}(1997)}]{Babel97}
{Babel}, J., \& {Montmerle}, T. 1997, \aap, 323, 121

\bibitem[{{Bernasconi} \& {Maeder}(1996)}]{Bernasconi96}
{Bernasconi}, P.~A., \& {Maeder}, A. 1996, \aap, 307, 829

\bibitem[{{Blondel} \& {Djie}(2006)}]{Blondel06}
{Blondel}, P.~F.~C., \& {Djie}, H.~R.~E.~T.~A. 2006, \aap, 456, 1045

\bibitem[{{B{\"o}hm} {et~al.}(2004){B{\"o}hm}, {Catala}, {Balona}, \&
  {Carter}}]{Bohm04}
{B{\"o}hm}, T., {Catala}, C., {Balona}, L., \& {Carter}, B. 2004, \aap, 427,
  907

\bibitem[{{B{\"o}hm} {et~al.}(2006){B{\"o}hm}, {Dupret.}, \&
  {Aynedjian}}]{Bohm06}
{B{\"o}hm}, T., {Dupret.}, M.~A., \& {Aynedjian}, H. 2006, Memorie della
  Societa Astronomica Italiana, 77, 362

\bibitem[{{Canizares} {et~al.}(2005){Canizares}, {Davis}, {Dewey}, {Flanagan},
  {Galton}, {Huenemoerder}, {Ishibashi}, {Markert}, {Marshall}, {McGuirk},
  {Schattenburg}, {Schulz}, {Smith}, \& {Wise}}]{hetg05}
{Canizares}, C.~R., {Davis}, J.~E., {Dewey}, D., {Flanagan}, K.~A., {Galton},
  E.~B., {Huenemoerder}, D.~P., {Ishibashi}, K., {Markert}, T.~H., {Marshall},
  H.~L., {McGuirk}, M., {Schattenburg}, M.~L., {Schulz}, N.~S., {Smith}, H.~I.,
  \& {Wise}, M. 2005, \pasp, 117, 1144

\bibitem[{{Catala} {et~al.}(2007){Catala}, {Alecian}, {Donati}, {Wade},
  {Landstreet}, {B{\"o}hm}, {Bouret}, {Bagnulo}, {Folsom}, \&
  {Silvester}}]{Catala07}
{Catala}, C., {Alecian}, E., {Donati}, J.-F., {Wade}, G.~A., {Landstreet},
  J.~D., {B{\"o}hm}, T., {Bouret}, J.-C., {Bagnulo}, S., {Folsom}, C., \&
  {Silvester}, J. 2007, \aap, 462, 293

\bibitem[{{Damiani} {et~al.}(1994){Damiani}, {Micela}, {Sciortino}, \&
  {Harnden}}]{Damiani94}
{Damiani}, F., {Micela}, G., {Sciortino}, S., \& {Harnden}, Jr., F.~R. 1994,
  \apj, 436, 807

\bibitem[{{D'Antona} \& {Mazzitelli}(1994)}]{Dantona94}
{D'Antona}, F., \& {Mazzitelli}, I. 1994, \apjs, 90, 467

\bibitem[{{Davis}(2001)}]{Davis01}
{Davis}, J.~E. 2001, \apj, 562, 575

\bibitem[{{Donati} {et~al.}(1997){Donati}, {Semel}, {Carter}, {Rees}, \&
  {Collier Cameron}}]{Donati97}
{Donati}, J.-F., {Semel}, M., {Carter}, B.~D., {Rees}, D.~E., \& {Collier
  Cameron}, A. 1997, \mnras, 291, 658

\bibitem[{{Drake} {et~al.}(2005){Drake}, {Testa}, \& {Hartmann}}]{Drake05}
{Drake}, J.~J., {Testa}, P., \& {Hartmann}, L. 2005, \apjl, 627, L149

\bibitem[{{Feigelson} {et~al.}(2003){Feigelson}, {Lawson}, \&
  {Garmire}}]{Feigelson03}
{Feigelson}, E.~D., {Lawson}, W.~A., \& {Garmire}, G.~P. 2003, \apj, 599, 1207

\bibitem[{{Flaccomio} {et~al.}(2005){Flaccomio}, {Micela}, {Sciortino},
  {Feigelson}, {Herbst}, {Favata}, {Harnden}, \& {Vrtilek}}]{Flaccomio05}
{Flaccomio}, E., {Micela}, G., {Sciortino}, S., {Feigelson}, E.~D., 
  {Herbst}, W., {Favata}, F., {Harnden}, Jr., F.~R., \& {Vrtilek}, S.~D. 
  2005, \apjs, 160, 450

\bibitem[{{Fruscione} {et~al.}(2006){Fruscione}, {McDowell}, {Allen},
  {Brickhouse}, {Burke}, {Davis}, {Durham}, {Elvis}, {Galle}, {Harris},
  {Huenemoerder}, {Houck}, {Ishibashi}, {Karovska}, {Nicastro}, {Noble},
  {Nowak}, {Primini}, {Siemiginowska}, {Smith}, \& {Wise}}]{CIAO}
{Fruscione}, A., {McDowell}, J.~C., {Allen}, G.~E., {Brickhouse}, N.~S.,
  {Burke}, D.~J., {Davis}, J.~E., {Durham}, N., {Elvis}, M., {Galle}, E.~C.,
  {Harris}, D.~E., {Huenemoerder}, D.~P., {Houck}, J.~C., {Ishibashi}, B.,
  {Karovska}, M., {Nicastro}, F., {Noble}, M.~S., {Nowak}, M.~A., {Primini},
  F.~A., {Siemiginowska}, A., {Smith}, R.~K., \& {Wise}, M. 2006, in Presented
  at the Society of Photo-Optical Instrumentation Engineers (SPIE) Conference,
  Vol. 6270, Observatory Operations: Strategies, Processes, and Systems. Edited
  by Silva, David R.; Doxsey, Rodger E.. Proceedings of the SPIE, Volume 6270,
  pp. 62701V (2006).

\bibitem[{{Gabriel} \& {Jordan}(1969)}]{Gabriel69}
{Gabriel}, A.~H., \& {Jordan}, C. 1969, \mnras, 145, 241

\bibitem[{{Garcia Lopez} {et~al.}(2006){Garcia Lopez}, {Natta}, {Testi}, \&
  {Habart}}]{GarciaLopez06}
{Garcia Lopez}, R., {Natta}, A., {Testi}, L., \& {Habart}, E. 2006, \aap, 459,
  837

\bibitem[{{Getman} {et~al.}(2005){Getman}, {Flaccomio}, {Broos}, {Grosso},
  {Tsujimoto}, {Townsley}, {Garmire}, {Kastner}, {Li}, {Harnden}, {Wolk},
  {Murray}, {Lada}, {Muench}, {McCaughrean}, {Meeus}, {Damiani}, {Micela},
  {Sciortino}, {Bally}, {Hillenbrand}, {Herbst}, {Preibisch}, \&
  {Feigelson}}]{Getman05}
{Getman}, K.~V., {Flaccomio}, E., {Broos}, P.~S., {Grosso}, N., {Tsujimoto},
  M., {Townsley}, L., {Garmire}, G.~P., {Kastner}, J., {Li}, J., {Harnden},
  Jr., F.~R., {Wolk}, S., {Murray}, S.~S., {Lada}, C.~J., {Muench}, A.~A.,
  {McCaughrean}, M.~J., {Meeus}, G., {Damiani}, F., {Micela}, G., {Sciortino},
  S., {Bally}, J., {Hillenbrand}, L.~A., {Herbst}, W., {Preibisch}, T., \&
  {Feigelson}, E.~D. 2005, \apjs, 160, 319

\bibitem[{{Gorenstein}(1975)}]{Gorenstein75}
{Gorenstein}, P. 1975, \apj, 198, 95

\bibitem[{{Grady} {et~al.}(1999){Grady}, {Woodgate}, {Bruhweiler}, {Boggess},
  {Plait}, {Lindler}, {Clampin}, \& {Kalas}}]{Grady99}
{Grady}, C.~A., {Woodgate}, B., {Bruhweiler}, F.~C., {Boggess}, A., {Plait},
  P., {Lindler}, D.~J., {Clampin}, M., \& {Kalas}, P. 1999, \apjl, 523, L151

\bibitem[{{Grady} {et~al.}(2004){Grady}, {Woodgate}, {Torres}, {Henning},
  {Apai}, {Rodmann}, {Wang}, {Stecklum}, {Linz}, {Williger}, {Brown},
  {Wilkinson}, {Harper}, {Herczeg}, {Danks}, {Vieira}, {Malumuth}, {Collins},
  \& {Hill}}]{Grady04}
{Grady}, C.~A., {Woodgate}, B., {Torres}, C.~A.~O., {Henning}, T., {Apai}, D.,
  {Rodmann}, J., {Wang}, H., {Stecklum}, B., {Linz}, H., {Williger}, G.~M.,
  {Brown}, A., {Wilkinson}, E., {Harper}, G.~M., {Herczeg}, G.~J., {Danks}, A.,
  {Vieira}, G.~L., {Malumuth}, E., {Collins}, N.~R., \& {Hill}, R.~S. 2004,
  \apj, 608, 809

\bibitem[{{Grady} {et~al.}(2007){Grady}, {Schneider}, {Hamaguchi}, K., 
  {Sitko}, {Carpenter}, {Hines}, {Collins}, {Williger}, {Woodgate}, {Henning}, 
  {M{\'e}nard}, {Wilner}, {Petre}, {Palunas}, {Quirrenbach}, {Nuth}, 
  {Silverstone}, {Kim}}]{Grady07}
{Grady}, C.~A., {Schneider}, G., {Hamaguchi}, K., {Sitko}, M.~L., 
  {Carpenter}, W.~J., {Hines}, D., {Collins}, K.~A., {Williger}, G.~M.,
  {Woodgate}, B.~E., {Henning}, T., {M{\'e}nard}, F., {Wilner}, D.,
  {Petre}, R., {Palunas}, P., {Quirrenbach}, A., {Nuth}, III, J.~A.,
  {Silverstone}, M.~D. \& {Kim}, J.~S. 2007, \apj, 665, 1391

\bibitem[{{Gregory} {et~al.}(2006){Gregory}, {Jardine}, {Cameron}, \& 
  {Donati}}]{Gregory06}
{Gregory}, S.~G., {Jardine}, M., {Cameron}, A.~C. \& {Donati}, J.-F. 2006,
  \mnras, 373,  827
  
\bibitem[{{Grevesse} \& {Sauval}(1998)}]{GrevesseSauval}
{Grevesse}, N., \& {Sauval}, A.~J. 1998, Space Science Reviews, 85, 161

\bibitem[{{G{\"u}del} {et~al.}(2007){G{\"u}del}, {Briggs}, {Arzner}, {Audard},
  {Bouvier}, {Feigelson}, {Franciosini}, {Glauser}, {Grosso}, {Micela},
  {Monin}, {Montmerle}, {Padgett}, {Palla}, {Pillitteri}, {Rebull}, {Scelsi},
  {Silva}, {Skinner}, {Stelzer}, \& {Telleschi}}]{Gudel07}
{G{\"u}del}, M., {Briggs}, K.~R., {Arzner}, K., {Audard}, M., {Bouvier}, J.,
  {Feigelson}, E.~D., {Franciosini}, E., {Glauser}, A., {Grosso}, N., {Micela},
  G., {Monin}, J.-L., {Montmerle}, T., {Padgett}, D.~L., {Palla}, F.,
  {Pillitteri}, I., {Rebull}, L., {Scelsi}, L., {Silva}, B., {Skinner}, S.~L.,
  {Stelzer}, B., \& {Telleschi}, A. 2007, \aap, 468, 353

\bibitem[{{Guimar{\~a}es} {et~al.}(2006){Guimar{\~a}es}, {Alencar}, {Corradi},
  \& {Vieira}}]{Guimaraes06}
{Guimar{\~a}es}, M.~M., {Alencar}, S.~H.~P., {Corradi}, W.~J.~B., \& {Vieira},
  S.~L.~A. 2006, \aap, 457, 581

\bibitem[{{G{\"u}nther} {et~al.}(2006){G{\"u}nther}, {Liefke}, {Schmitt},
  {Robrade}, \& {Ness}}]{Gunther06}
{G{\"u}nther}, H.~M., {Liefke}, C., {Schmitt}, J.~H.~M.~M., {Robrade}, J., \&
  {Ness}, J.-U. 2006, \aap, 459, L29

\bibitem[{{Hamaguchi} {et~al.}(2005){Hamaguchi}, {Yamauchi}, \&
  {Koyama}}]{Hamaguchi05}
{Hamaguchi}, K., {Yamauchi}, S., \& {Koyama}, K. 2005, \apj, 618, 360

\bibitem[{{Herbig}(1960)}]{Herbig60}
{Herbig}, G.~H. 1960, \apjs, 4, 337

\bibitem[{{Houck} \& {Denicola}(2000)}]{Houck00}
{Houck}, J.~C., \& {Denicola}, L.~A. 2000, in ASP Conf. Ser. 216: Astronomical
  Data Analysis Software and Systems IX, Vol.~9, 591

\bibitem[{{Hu} {et~al.}(1991){Hu}, {Blondel}, {The}, {Tjin A Djie}, {de
  Winter}, {Catala}, \& {Talavera}}]{Hu91}
{Hu}, J.~Y., {Blondel}, P.~F.~C., {The}, P.~S., {Tjin A Djie}, H.~R.~E., {de
  Winter}, D., {Catala}, C., \& {Talavera}, A. 1991, \aap, 248, 150

\bibitem[{{Hu} {et~al.}(1989){Hu}, {The}, \& {de Winter}}]{Hu89}
{Hu}, J.~Y., {The}, P.~S., \& {de Winter}, D. 1989, \aap, 208, 213

\bibitem[{{Hubrig} {et~al.}(2007){Hubrig}, {Pogodin}, {Yudin}, {Sch{\"o}ller},
  \& {Schnerr}}]{Hubrig07}
{Hubrig}, S., {Pogodin}, M.~A., {Yudin}, R.~V., {Sch{\"o}ller}, M., \&
  {Schnerr}, R.~S. 2007, \aap, 463, 1039

\bibitem[{{Hubrig} {et~al.}(2004){Hubrig}, {Sch{\"o}ller}, \&
  {Yudin}}]{Hubrig04}
{Hubrig}, S., {Sch{\"o}ller}, M., \& {Yudin}, R.~V. 2004, \aap, 428, L1

\bibitem[{{Huenemoerder} {et~al.}(2007){Huenemoerder}, {Kastner}, {Testa},
  {Schulz}, \& {Weintraub}}]{Huenemoerder07}
{Huenemoerder}, D.~P., {Kastner}, J.~H., {Testa}, P., {Schulz}, N.~S., \&
  {Weintraub}, D.~A. 2007, \apj, 671, 000

\bibitem[{{Huenemoerder} {et~al.}(2006){Huenemoerder}, {Testa}, \&
  {Buzasi}}]{Huenemoerder06}
{Huenemoerder}, D.~P., {Testa}, P., \& {Buzasi}, D.~L. 2006, \apj, 650, 1119

\bibitem[{{Ishibashi} {et~al.}(2006){Ishibashi}, {Dewey}, {Huenemoerder}, \&
  {Testa}}]{Ishibashi06}
{Ishibashi}, K., {Dewey}, D., {Huenemoerder}, D.~P., \& {Testa}, P. 2006,
  \apjl, 644, L117

\bibitem[{{Kastner} {et~al.}(2002){Kastner}, {Huenemoerder}, {Schulz},
  {Canizares}, \& {Weintraub}}]{Kastner02}
{Kastner}, J.~H., {Huenemoerder}, D.~P., {Schulz}, N.~S., {Canizares}, C.~R.,
  \& {Weintraub}, D.~A. 2002, \apj, 567, 434

\bibitem[{{Larson}(1972)}]{Larson72}
{Larson}, R.~B. 1972, \mnras, 157, 121

\bibitem[{{Luhman}(2004)}]{Luhman04}
{Luhman}, K.~L. 2004, \apj, 616, 1033

\bibitem[{{Maggio} {et~al.}(2007){Maggio}, {Flaccomio}, {Favata},
  {Micela}, {Sciortino}, {Feigelson}, \& {Getman}}]{Maggio07}
{Maggio}, A., {Flaccomio}, E., {Favata}, F., {Micela}, G., 
  {Sciortino}, S., {Feigelson}, E.~D. \& {Getman}, K.~V. 2007, 
  \apj, 660, 1462
   
\bibitem[{{Malfait} {et~al.}(1998){Malfait}, {Bogaert}, \&
  {Waelkens}}]{Malfait98}
{Malfait}, K., {Bogaert}, E., \& {Waelkens}, C. 1998, \aap, 331, 211

\bibitem[{{Mannings} \& {Sargent}(1997)}]{Mannings97}
{Mannings}, V., \& {Sargent}, A.~I. 1997, \apj, 490, 792

\bibitem[{{Muzerolle} {et~al.}(2004){Muzerolle}, {D'Alessio}, {Calvet}, \&
  {Hartmann}}]{Muzerolle04}
{Muzerolle}, J., {D'Alessio}, P., {Calvet}, N., \& {Hartmann}, L. 2004, \apj,
  617, 406

\bibitem[{{Ness} {et~al.}(2004){Ness}, {G{\"u}del}, {Schmitt}, {Audard}, \&
  {Telleschi}}]{Ness04}
{Ness}, J.-U., {G{\"u}del}, M., {Schmitt}, J.~H.~M.~M., {Audard}, M., \&
  {Telleschi}, A. 2004, \aap, 427, 667

\bibitem[{{Palla} \& {Stahler}(1993)}]{Palla93}
{Palla}, F., \& {Stahler}, S.~W. 1993, \apj, 418, 414

\bibitem[{{Perryman} {et~al.}(1997){Perryman}, {Lindegren}, {Kovalevsky},
  {Hoeg}, {Bastian}, {Bernacca}, {Cr{\'e}z{\'e}}, {Donati}, {Grenon}, {van
  Leeuwen}, {van der Marel}, {Mignard}, {Murray}, {Le Poole}, {Schrijver},
  {Turon}, {Arenou}, {Froeschl{\'e}}, \& {Petersen}}]{Perryman97}
{Perryman}, M.~A.~C., {Lindegren}, L., {Kovalevsky}, J., {Hoeg}, E., {Bastian},
  U., {Bernacca}, P.~L., {Cr{\'e}z{\'e}}, M., {Donati}, F., {Grenon}, M., {van
  Leeuwen}, F., {van der Marel}, H., {Mignard}, F., {Murray}, C.~A., {Le
  Poole}, R.~S., {Schrijver}, H., {Turon}, C., {Arenou}, F., {Froeschl{\'e}},
  M., \& {Petersen}, C.~S. 1997, \aap, 323, L49

\bibitem[{{Preibisch} {et~al.}(2005){Preibisch}, {Kim}, {Favata}, {Feigelson},
  {Flaccomio}, {Getman}, {Micela}, {Sciortino}, {Stassun}, {Stelzer}, \&
  {Zinnecker}}]{Preibisch05}
{Preibisch}, T., {Kim}, Y.-C., {Favata}, F., {Feigelson}, E.~D., {Flaccomio},
  E., {Getman}, K., {Micela}, G., {Sciortino}, S., {Stassun}, K., {Stelzer},
  B., \& {Zinnecker}, H. 2005, \apjs, 160, 401

\bibitem[{{Sanz-Forcada} {et~al.}(2003){Sanz-Forcada}, {Brickhouse}, \&
  {Dupree}}]{Sanz03a}
{Sanz-Forcada}, J., {Brickhouse}, N.~S., \& {Dupree}, A.~K. 2003, \apjs, 145,
  147

\bibitem[{{Schmitt} {et~al.}(2005){Schmitt}, {Robrade}, {Ness}, {Favata}, \&
  {Stelzer}}]{Schmitt05}
{Schmitt}, J.~H.~M.~M., {Robrade}, J., {Ness}, J.-U., {Favata}, F., \&
  {Stelzer}, B. 2005, \aap, 432, L35

\bibitem[{{Skinner} {et~al.}(2004){Skinner}, {G{\"u}del}, {Audard}, \&
  {Smith}}]{Skinner04}
{Skinner}, S.~L., {G{\"u}del}, M., {Audard}, M., \& {Smith}, K. 2004, \apj,
  614, 221

\bibitem[{{Skinner} \& {Yamauchi}(1996)}]{Skinner96}
{Skinner}, S.~L., \& {Yamauchi}, S. 1996, \apj, 471, 987

\bibitem[{{Smith} {et~al.}(2001){Smith}, {Brickhouse}, {Liedahl}, \&
  {Raymond}}]{Smith01}
{Smith}, R.~K., {Brickhouse}, N.~S., {Liedahl}, D.~A., \& {Raymond}, J.~C.
  2001, \apjl, 556, L91

\bibitem[{{Stelzer} {et~al.}(2006){Stelzer}, {Micela}, {Hamaguchi}, \&
  {Schmitt}}]{Stelzer06}
{Stelzer}, B., {Micela}, G., {Hamaguchi}, K., \& {Schmitt}, J.~H.~M.~M. 2006,
  \aap, 457, 223

\bibitem[{{Stelzer} \& {Schmitt}(2004)}]{Stelzer04}
{Stelzer}, B., \& {Schmitt}, J.~H.~M.~M. 2004, \aap, 418, 687

\bibitem[{{Swartz} {et~al.}(2005){Swartz}, {Drake}, {Elsner}, {Ghosh}, {Grady},
  {Wassell}, {Woodgate}, \& {Kimble}}]{Swartz05}
{Swartz}, D.~A., {Drake}, J.~J., {Elsner}, R.~F., {Ghosh}, K.~K., {Grady},
  C.~A., {Wassell}, E., {Woodgate}, B.~E., \& {Kimble}, R.~A. 2005, \apj, 628,
  811

\bibitem[{{Tatulli} {et~al.}(2007){Tatulli}, {Isella}, {Natta}, {Testi},
  {Marconi}, {Malbet}, {Stee}, {Petrov}, \& {et al.}}]{Tatulli07}
{Tatulli}, E., {Isella}, A., {Natta}, A., {Testi}, L., {Marconi}, A., {Malbet},
  F., {Stee}, P., {Petrov}, R.~G., \& {et al.} 2007, \aap, 464, 55

\bibitem[{{Telleschi} {et~al.}(2007){Telleschi}, {G{\"u}del}, {Briggs},
  {Skinner}, {Audard}, \& {Franciosini}}]{Telleschi07}
{Telleschi}, A., {G{\"u}del}, M., {Briggs}, K.~R., {Skinner}, S.~L., {Audard},
  M., \& {Franciosini}, E. 2007, \aap, 468, 541

\bibitem[{{Telleschi} {et~al.}(2005){Telleschi}, {G{\"u}del}, {Briggs},
  {Audard}, {Ness}, \& {Skinner}}]{Telleschi05}
{Telleschi}, A., {G{\"u}del}, M., {Briggs}, K.~R., {Audard}, M.,
  {Ness}, J.~U., \& {Skinner}, S.~L. 2005, \apj, 622, 653

\bibitem[{{Testa} {et~al.}(2004){Testa}, {Drake}, \& {Peres}}]{Testa04b}
{Testa}, P., {Drake}, J., \& {Peres}, G. 2004, \apj, 617, 508

\bibitem[{{Tout} \& {Pringle}(1995)}]{Tout95}
{Tout}, C.~A., \& {Pringle}, J.~E. 1995, \mnras, 272, 528

\bibitem[{{van den Ancker} {et~al.}(1998){van den Ancker}, {de Winter}, \&
  {Tjin A Djie}}]{vdAncker98}
{van den Ancker}, M.~E., {de Winter}, D., \& {Tjin A Djie}, H.~R.~E. 1998,
  \aap, 330, 145

\bibitem[{{Wade} {et~al.}(2007){Wade}, {Bagnulo}, {Drouin}, {Landstreet}, \&
  {Monin}}]{Wade07}
{Wade}, G.~A., {Bagnulo}, S., {Drouin}, D., {Landstreet}, J.~D., \& {Monin}, D.
  2007, \mnras, 376, 1145

\bibitem[{{Wade} {et~al.}(2005){Wade}, {Drouin}, {Bagnulo}, {Landstreet},
  {Mason}, {Silvester}, {Alecian}, {B{\"o}hm}, {Bouret}, {Catala}, \&
  {Donati}}]{Wade05}
{Wade}, G.~A., {Drouin}, D., {Bagnulo}, S., {Landstreet}, J.~D., {Mason}, E.,
  {Silvester}, J., {Alecian}, E., {B{\"o}hm}, T., {Bouret}, J.-C., {Catala},
  C., \& {Donati}, J.-F. 2005, \aap, 442, L31

\bibitem[{{Walker} \& {Wolstencroft}(1988)}]{Walker88}
{Walker}, H.~J., \& {Wolstencroft}, R.~D. 1988, \pasp, 100, 1509

\bibitem[{{Waters} \& {Waelkens}(1998)}]{Waters98}
{Waters}, L.~B.~F.~M., \& {Waelkens}, C. 1998, \araa, 36, 233

\bibitem[{{Zinnecker} \& {Preibisch}(1994)}]{Zinnecker94}
{Zinnecker}, H., \& {Preibisch}, T. 1994, \aap, 292, 152

\end{thebibliography}

\end{document}